\documentclass[notitlepage,floatfix,superscriptaddress,nofootinbib,tightenlines,preprintnumbers,twocolumn]{revtex4-2}
\usepackage{amsmath,verbatim,latexsym,amssymb,indentfirst,mathrsfs,mathtools,amsthm,bbm,bm,hyperref,url,cancel,subcaption,enumitem}
\usepackage[font=small,labelfont=bf,format=plain,justification=raggedright,singlelinecheck=false]{caption}
\usepackage{verbatim,indentfirst}
\usepackage{siunitx}
\usepackage[title,titletoc]{appendix}
\hypersetup{nolinks=true}

\usepackage{graphicx}
\usepackage{epstopdf}

\usepackage{subcaption}

\renewcommand{\thesection}{\Roman{section}}
\renewcommand{\thesubsection}{\Roman{section} \Alph{subsection}}
\renewcommand{\thesubsubsection}{\Roman{section} \Alph{subsection} \arabic{subsubsection}}
\makeatletter
\def\p@subsection{}
\makeatother
\makeatletter
\def\p@subsubsection{}
\makeatother

\makeatletter
\newcommand\footnoteref[1]{\protected@xdef\@thefnmark{\ref{#1}}\@footnotemark}
\makeatother

\usepackage{soul}

\usepackage{color}
\usepackage[dvipsnames]{xcolor}
\usepackage[normalem]{ulem}

\newcommand{\MSS}{{\rm MSS}}
\newcommand{\GOE}{{\rm GOE}}

\newcommand{\tot}{ {\rm tot} }
\def\const{ {\rm const.} }   
\def\id{\mathbbm{1}}   

\newcommand{\Dim}{d}   
\newcommand{\0}{ {(0)} }
\newcommand{\1}{ {(1)} }
\newcommand{\2}{ {(2)} }

\newcommand{\JParen}{ {(j)} }
\newcommand{\KParen}{ {(k)} }

	\definecolor{blue(pigment)}{rgb}{0.2, 0.2, 0.6}
 
\renewcommand\th{ {\rm th} }

\newcommand*{\bra}[1]{\langle #1\rvert}
\newcommand*{\ket}[1]{\lvert #1 \rangle}
\newcommand*{\braket}[2]{\langle #1 \lvert #2 \rangle}
\newcommand*{\ketbra}[2]{\lvert #1 \rangle\!\langle #2 \rvert}

\hypersetup{final}

\begin{document}

\title{Numerical evidence for the non-Abelian eigenstate thermalization hypothesis}
%

%

\author{Aleksander~Lasek}
\email{alasek@umd.edu}
\affiliation{Joint Center for Quantum Information and Computer Science, NIST and University of Maryland, College Park, Maryland 20742, USA}
%

\author{Jae~Dong~Noh}
\affiliation{Department of Physics, University of Seoul, Seoul 02504, Korea}

\author{Jade~LeSchack}
\affiliation{Joint Center for Quantum Information and Computer Science, NIST and University of Maryland, College Park, Maryland 20742, USA}
\author{Nicole~Yunger~Halpern}
\affiliation{Joint Center for Quantum Information and Computer Science, NIST and University of Maryland, College Park, Maryland 20742, USA}
\affiliation{Institute for Physical Science and Technology, University of Maryland, College Park, MD 20742, USA}

\date{\today}

%
%
\begin{abstract} 

The eigenstate thermalization hypothesis (ETH) explains how generic quantum many-body systems thermalize internally. It implies that local operators' time-averaged expectation values approximately equal their thermal expectation values, regardless of microscopic details. The ETH's range of applicability therefore impacts theory and experiments. Murthy \emph{et al.} recently showed that non-Abelian symmetries conflict with the ETH. Such symmetries have excited interest in quantum thermodynamics lately, as they are equivalent to conserved quantities that fail to commute with each other and noncommutation is a quintessentially quantum phenomenon. Murthy \emph{et al.} proposed a non-Abelian ETH, which we support numerically. The numerics model a one-dimensional (1D) next-nearest-neighbor Heisenberg chain of 18 qubits. We represent local operators with matrices relative to an energy eigenbasis. The matrices bear out seven predictions of the non-Abelian ETH. We also prove analytically that the non-Abelian ETH exhibits a self-consistency property. The proof relies on a thermodynamic-entropy definition different from that in Murthy \emph{et al.} This work initiates the observation and application of the non-Abelian ETH.
\end{abstract}

{\let\newpage\relax\maketitle}

%
%
%
\section{Introduction}

\emph{Prima facie}, an isolated quantum system challenges the second law of thermodynamics. Such a system evolves unitarily; so its state's von Neumann entropy, for example, never changes. Yet generic quantum many-body systems thermalize internally: local operators' time-expectation values approximately equal thermal expectation values. 
This thermalization often arises from the eigenstate thermalization hypothesis (ETH)~\cite{DeutschThermalization1991,SrednickiThermalization1994,RigolThermalization2008}. To introduce the ETH, we denote by $A$ a local operator and by $H$ a nonintegrable many-body Hamiltonian. A matrix represents $A$ relative to the $H$ eigenbasis. The ETH posits a particular structure, reviewed below, for that matrix. Under three necessary conditions, the ETH implies that $A$ thermalizes in the above sense: (i) $A$ and $H$ obey the ETH. (ii) $H$ lacks degeneracies. (iii) The system begins in a microcanonical subspace, an eigenspace shared by the conserved quantities (\emph{charges}).\footnote{
More precisely, the eigenspace is shared by the charges other than $H$, which begins with a small variance. $H$ requires an exception because it generates the time evolution.}
The ETH has influenced many fields, including condensed-matter physics; quantum chaos; atomic, molecular, and optical (AMO) physics; quantum thermodynamics; and high-energy and nuclear physics~\cite{Gogolin_2016,DAlessio_16_From}.

Despite this widespread use, non-Abelian symmetries invalidate the ETH and its implication for thermalization. Murthy \emph{et al.} recently identified three conflicts between the ETH-based thermalization argument and a Hamiltonian $H$ given a non-Abelian symmetry~\cite{MurthyNAETH}. (We focus on continuous symmetries, equivalent to charge conservation~\cite{Majidy_23_Noncommuting}.)
First, $H$ has degeneracies, in contrast with condition (ii). Second, microcanonical subspaces do not necessarily exist at all or in abundance: due to the non-Abelian symmetry, $H$ conserves charges that fail to commute with each other. They do not share any eigenbasis and so do not necessarily share eigenspaces. Third, the ETH conflicts with the Wigner--Eckart theorem, a fundamental quantum result that dictates transition frequencies in AMO physics~\cite{shankar2008principles}. Hence noncommuting charges undermine expectations about quantum many-body thermalization.

Noncommuting charges have recently gained prominence in quantum thermodynamics~\cite{Majidy_23_Noncommuting}. Many thermodynamic setups conserve energy, particles or the total spin's $z$-component, and other quantities that commute with each other. For decades, charges were implicitly assumed to commute in multiple thermodynamic arguments; examples include derivations of the thermal state's form~\cite{Halpern_2018_HeatBaths2,guryanova_2016_thermodynamics,Lostaglio_2017_MaxEnt,Halpern_2016_microcanonical} and of the Onsager relations~\cite{Manzano_22_Non}. This implicit assumption became widely recognized only a decade ago~\cite{lostaglio_2014_thermodynamics,Halpern_2018_HeatBaths2}. Yet noncommutation is a quintessential quantum phenomenon, enabling measurement disturbance, uncertainty relations, and quantum error correction. Therefore, a growing subfield of quantum thermodynamics centers on the question \emph{how does thermodynamics change when charges fail to commute?}~\cite{Majidy_23_Noncommuting}. Altered thermodynamic phenomena include thermodynamic-entropy production~\cite{Manzano_22_Non,Upadhyaya_24_Non}, bipartite entanglement~\cite{Majidy_23_Non}, quantum-computational universality~\cite{Marvian_23_Non}, and measurement-induced phases~\cite{Majidy_23_Critical}, as well as the ETH~\cite{MurthyNAETH}.

Murthy \emph{et al.} proposed a \emph{non-Abelian ETH} for Hamiltonians that conserve noncommuting charges~\cite{MurthyNAETH}. The authors also began uncovering the non-Abelian ETH's implications for thermalization, studying an $N$-qubit system subject to SU(2) symmetry. Under certain conditions, local operators' time-averaged expectation values equal their thermal expectation values, to within corrections of $O(N^{-1})$---the same size as in the absence of noncommuting charges. The corrections are polynomially larger---of $O(N^{-1/2})$---under other conditions. The system may remain somewhat farther from its thermal fixed point than usual (than in the absence of any non-Abelian symmetry). Thermalization dissipates information about initial conditions, so this anomaly might support information retention in quantum memories~\cite{MurthyNAETH,25_Noh_Kubo}. This anomalous thermalization requires numerical and experimental support. 

Noh recently initiated the numerical investigation of the non-Abelian ETH~\cite{JaeDongXXZ}. He studied a two-dimensional (2D) Heisenberg model on a rectangular lattice. 
The operators $A$ were combinations of products of two Pauli operators. Noh represented those operators as matrices relative to an energy eigenbasis. The matrix elements followed trends dependent on total spin quantum number. Also, the matrix elements exhibited variances that obeyed a prediction of the non-Abelian ETH.
The analysis did not involve the Wigner--Eckart theorem, which the non-Abelian ETH does. Additionally, the analysis focused on a subspace defined in terms of not only the SU(2) symmetry, but also extra symmetries.

We progress beyond Noh's analysis, providing extensive numerical evidence for the non-Abelian ETH. We model a 1D chain of 18 qubits, which couple via nearest-neighbor and next-nearest-neighbor Heisenberg interactions. The Hamiltonian exhibits SU(2) symmetry (and no extraneous symmetries): it conserves the total spin's $x$- $y$- and $z$-components, which fail to commute with each other. 
The Hamiltonian and local operators, we demonstrate, exhibit seven properties
predicted by the non-Abelian ETH. This work provides the first thorough numerical support for the non-Abelian ETH.

Additionally, we demonstrate analytically that the non-Abelian ETH exhibits a self-consistency property introduced by Srednicki~\cite{Srednicki_96_Thermal}. This conclusion holds if the non-Abelian ETH depends on a thermodynamic entropy smaller than the one posited in~\cite{MurthyNAETH}. The reason stems from the Wigner--Eckart theorem. Just as the theorem implies selection rules in AMO physics (atoms cannot undergo certain state transitions), the theorem dictates that certain energy eigenstates cannot contribute to the thermodynamic entropy.

The rest of this paper is organized as follows. Section~\ref{sec_background} introduces background information. We argue in Sec.~\ref{sec_2s_main} for the non-Abelian ETH's self-consistency. Section~\ref{sec_Model} introduces our model and computational methods. Section~\ref{sec_results} contains our main results: numerical evidence for the non-Abelian ETH. Section~\ref{sec_Conclusions} details opportunities for future work.

\section{Technical background}
\label{sec_background}

This section relates several types of background information. In Sec.~\ref{sec_Backgrnd_ETH}, we review the ETH. We review spherical tensor operators in Sec.~\ref{sec_Backgrnd_Oprs}; they form a convenient basis, in SU(2)-symmetric settings, for the space of operators $A$. Spherical tensor operators obey the Wigner–Eckart theorem, also reviewed in this subsection. Section~\ref{sec_Backgrnd_NAETH} introduces the non-Abelian ETH.

\subsection{ETH}
\label{sec_Backgrnd_ETH}

The ETH governs a generic quantum many-body system of the following type. The Hamiltonian is nonintegrable, lacks non-Abelian symmetries, and has eigenvalues $E_\alpha$ associated with eigenstates $\ket{\alpha}$. Let $A$ denote an operator defined on the system's Hilbert space. We can represent $A$ as a matrix, relative to the energy eigenbasis, with elements $\bra{\alpha} A \ket{\alpha'}$. 
Most local operators $A$ are expected to obey ETH, which depends on the following quantities. Define the average energy $\mathcal{E} \coloneqq (E_\alpha + E_{\alpha'})/2$ and the difference $\omega  \coloneqq  E_\alpha - E_{\alpha'}$.
The smooth, real function $\mathcal{A}(\mathcal{E})$ denotes the microcanonical average of $A$. If $A$ acts nontrivially on just $O(N^0) \equiv O(1)$ particles, then $\mathcal{A}$ is $O(1)$.\footnote{
We use big-$O$ notation to mean \emph{scales as}, as in many-body physics. The notation has a more precise significance in quantum information theory.}
Also $f ( \mathcal{E}, \omega )$ denotes a smooth, real, $O(1)$ function.
The thermodynamic entropy $S_\th (\mathcal{E})$ equals the logarithm of the density of states (DOS). All logarithms in this paper are base-$e$. The matrix $R$ consists of erratically fluctuating $O(1)$ numbers~\cite{Foini_19_Eigenstate,Jorge_2022_ETHprobability,Wang_2022_ETH_RMT}. 
$H$ and $A$ obey the ETH if~\cite{DeutschThermalization1991,SrednickiThermalization1994,RigolThermalization2008}
\begin{align}
   \label{eq_ETH}
   \bra{\alpha} A \ket{\alpha'}
   =  \mathcal{A} (\mathcal{E}) \, \delta_{\alpha \alpha'}
   +  e^{ - S_\th ( \mathcal{E} ) / 2}  \,
   f ( \mathcal{E}, \omega )  \,
   R_{\alpha \alpha'} \, .
\end{align}

The two terms have different physical significances.
The first, \emph{diagonal} term assumes nonzero values only if $\bra{\alpha} A \ket{\alpha'}$ lies on the matrix diagonal. The diagonal term dictates the time-averaged expectation value of $A$. The final, \emph{off-diagonal} term can assume nonzero values also off the matrix diagonal. This term governs fluctuations, across time, of the local operator's instantaneous expectation value.

\subsection{Spherical tensor operators and Wigner–Eckart theorem}
\label{sec_Backgrnd_Oprs}

Spherical tensor operators form a mathematical toolkit for analyzing operators $A$ in SU(2)-symmetric settings. Throughout the rest of this paper, we focus on the following setup, refined in Sec.~\ref{sec_Model}, unless we state otherwise. Consider a quantum many-body system governed by an SU(2)-symmetric Hamiltonian $H$.\footnote{
Results overviewed in the rest of this section are expected to generalize to other non-Abelian symmetries~\cite{MurthyNAETH}.}
The system consists of $N  \gg  1$ qubits. Denote the global spin components\footnote{
We invoke spin due to its prevalence in quantum information processing. However, any other angular momentum can replace spin without changing our arguments.}
by $S_{a = x, y, z}$ and the global spin-squared operator by $\vec{S}^2$.

Due to the SU(2) symmetry, $H$ shares an eigenbasis 
$\{ \ket{\alpha, m} \}$ with $\vec{S}^2$ and $S_z$. The $\alpha$ labels the eigenenergy $E_\alpha$ and (total) spin quantum number $s_\alpha$.\footnote{
$\{ \alpha \}$ maps bijectively to $\{ E_\alpha \}$ and surjectively onto $\{ s_\alpha \}$. The discrepancy arises because $s_\alpha$ assumes only $O(N)$ values, whereas $E_\alpha$ assumes $2^N$ values (some of which equal each other).}
$m$ denotes the (total spin) magnetic quantum number. We set $\hbar$ to 1:
\begin{align}
   \label{eq_Eigenthings}
   & H \ket{\alpha, m}  
   =  E_\alpha  \ket{\alpha, m} , \\
   & \vec{S}^2  \ket{\alpha, m}  
   =  s_\alpha (s_\alpha + 1)  \ket{\alpha, m} ,
   \quad \text{and} \\
   & S_z  \ket{\alpha, m}  
   =  m  \ket{\alpha, m} .
\end{align}
The raising and lowering operators 
$S_\pm  \coloneqq  S_x  \pm  i S_y$ obey the commutation relations
$[S_z, S_{\pm}] = \pm S_{\pm}$ and $[S_+, S_-] = 2 S_z$.

The \emph{spherical tensor operators} are tensors formed from operator components $T^\KParen_q$ \cite{shankar2008principles}.
For conciseness, we call the $T^\KParen_q$s themselves spherical tensor operators. The $T^\KParen_q$s span the space of operators defined on $\mathbb{C}^{2N}$. Therefore, we will focus on spherical-tensor-operator $A$s without loss of generality. The \emph{rank} $k$ resembles $s_\alpha$; and $q = -k, k+1, \ldots, k$ resembles $m$. When commuted with spin operators, the $T^\KParen_q$s transform similarly to eigenstates $\ket{\alpha, m}$ acted on by spin operators:
\begin{align}
   \label{eq_Def_Tensors}
   & \left[ S_z ,  T^\KParen_q \right]
   =  q \,  T^\KParen_q ,
   \quad \text{and} \\
   & \left[ S_\pm ,  T^\KParen_q  \right]
   =  \sqrt{ ( k \mp q) ( k \pm q + 1) }  \:
   T^\KParen_{q \pm 1} \, .
\end{align}
These equations imply that $T^\KParen_q$ transforms irreducibly under SU(2); SU(2) elements cannot change $k$.

A simple example offers intuition about spherical tensor operators. Suppose that an atom 
absorbs a photon. (We temporarily break our focus on many-body systems, for simplicity.) A spherical tensor operator represents the photon's action on the atom's state. Since the photon has a spin quantum number of 1, the operator has a rank $k = 1$. If the atom gains/loses a quantum of $z$-type spin angular momentum, then
$S_\pm  \propto  T^\1_\pm$ acts on the atom's state.

We extend this example to motivate the Wigner–Eckart theorem. Suppose that the atom began in a state $\ket{\alpha', m'}$. Consider measuring the atom's energy, $\vec{S}^2$, and and $S_z$ after the photon absorption. With what probability do the outcomes $E_\alpha$, $s_\alpha$, and $m$ obtain---or, rather, with what probability amplitude? All such probability amplitudes (in and beyond the atom example) have forms dictated by the \emph{Wigner–Eckart theorem} \cite{shankar2008principles}, 
\begin{align}
   \label{eq_WE_thm} &
   \bra{\alpha, m}  T^\KParen_q  \ket{\alpha', m'}
   \\ \nonumber &
   =  \langle s_\alpha \, , m | s_{\alpha'} \, , m' \, ; k, q \rangle
   \langle \alpha || T^\KParen || \alpha' \rangle .
\end{align}

The first factor on the right-hand side (RHS) is a \emph{Clebsch–Gordan coefficient}. It serves as a conversion factor between (i) an eigenstate $\ket{\alpha \, , m}$ shared by the global operators $\vec{S}^2$ and $S_z$ and (ii) a tensor product (of $\ket{{\alpha'} \, , m'}$ and a state labeled by quantum numbers $k$ and $q$). The final factor in Eq.~\eqref{eq_WE_thm} is a \emph{reduced matrix element}. It follows from dividing the Clebsch–Gordan coefficient out from the probability amplitude. $\langle \alpha || T^\KParen || \alpha' \rangle$ does not depend on any magnetic-type quantum number ($m$, $m'$, or $q$).

The Wigner–Eckart theorem clarifies why non-Abelian symmetry invalidates the ETH. The ETH purports to dictate matrix elements' forms, but the theorem dictates them provably. Those forms depend on the SU(2) symmetry, according to the theorem, but the ETH does not encode that symmetry. In summary, the ETH clashes with the Wigner–Eckart theorem in attempting to stipulate matrix elements' structures~\cite{MurthyNAETH,Majidy_23_Noncommuting}.

\subsection{Non-Abelian ETH}
\label{sec_Backgrnd_NAETH}

The non-Abelian ETH is an ansatz for the forms of reduced matrix elements 
$\langle \alpha || T^\KParen || \alpha' \rangle$.
Throughout the rest of this paper, we focus on $T^\KParen_q$s that operate nontrivially on just $O(1)$ qubits: such local operators are expected to obey ETHs. To introduce the non-Abelian ETH, we define the average 
$\mathcal{E}  \coloneqq  (E_\alpha + E_{\alpha'}) / 2$ 
and difference
$\omega  \coloneqq  E_\alpha - E_{\alpha'}$ 
as in Sec.~\ref{sec_Backgrnd_ETH}.
Similarly, we define the average
$\mathcal{S}  \coloneqq  (s_\alpha + s_{\alpha'})/2$
and difference
$\nu  \coloneqq  s_\alpha  -  s_{\alpha'}$
of spin quantum numbers.
$\mathcal{T}^\KParen$ and $f^{(T)}_{ \nu }$ denote smooth $O(1)$ functions.\footnote{
Each function is expected to be smooth in $\mathcal{E}/N$ and $\mathcal{S}/N$ in the thermodynamic limit. We invoke densities because $E_\alpha$ and $s_\alpha$ are quantized. If $N$ is finite, $\mathcal{E}$ and $\mathcal{S}$ are discrete. $\mathcal{T}^\KParen$ and $f^{(T)}_{ \nu }$ cannot be smooth functions of discrete arguments. However, each function can be smooth in each density in the thermodynamic limit. The functions' analogues, in the conventional ETH, are similarly smooth in $\mathcal{E}/N$.}
$R^{(T)}_{\alpha \alpha'}$ denotes a matrix of erratically fluctuating numbers. The ordinary ETH's analogous matrix elements, analyzed individually, obey Gaussian distributions \cite{DAlessio_16_From}. One may expect the non-Abelian ETH's matrix elements to obey Gaussian distributions similarly. Our numerics, below, support this expectation.
An operator $T^\KParen_q$ and a Hamiltonian $H$ obey the non-Abelian ETH if~\cite{MurthyNAETH}
\begin{align}
   \label{eq_NAETH}
   \langle \alpha || T^\KParen || \alpha' \rangle
   & =  \mathcal{T}^\KParen \left(  \mathcal{E} \, , \mathcal{S}  \right)  \,  \delta_{\alpha \alpha'} 
   \\  \nonumber & \quad
   +  e^{- S_\th \left( \mathcal{E} \, , \, \mathcal{S}  \right) / 2}  \, 
   f^{(T)}_{ \nu }  
      \left(  \mathcal{E}  \, ,  \mathcal{S} \, , \omega  \right)  \,
   R^{(T)}_{\alpha \alpha'} \, .
\end{align}
The first term is nonzero only along the matrix's \emph{block} diagonal.\footnote{
Relative to $\{ \ket{\alpha, m} \}$, $T^\KParen_q$ is represented by a block-diagonal matrix. Each block corresponds to one $\alpha$ value. Even if $m \neq m'$,
$\bra{\alpha, m} T^\KParen_q \ket{\alpha, m'}$ lies on the block diagonal. By the Wigner--Eckart theorem, the reduced matrix element 
$\langle \alpha || T^\KParen || \alpha \rangle$ does not depend on $m$ or $m'$. Therefore, 
$\langle \alpha || T^\KParen || \alpha \rangle$
appears in all the matrix elements in the diagonal block.}
(We call reduced matrix elements \emph{block-diagonal} if $\alpha'=\alpha$ and call them \emph{off–block-diagonal} otherwise.).

In~\cite{MurthyNAETH}, $S_{\th}(\mathcal{E},\mathcal{S})$ denoted the thermodynamic
entropy at the energy $\mathcal{E}$ and spin quantum number $\mathcal{S}$. We define $S_{\th}(\mathcal{E},\mathcal{S})$ as follows. Begin with the thermodynamic entropy at $\mathcal{E}$ and $\mathcal{S}$. Subtract off $\textrm{ln} (2\mathcal{S}+1)$. (The subtraction eliminates from $S_\th$ any dependence on any degeneracy attributable to the magnetic spin quantum number.~\footnote{
The magnetic-spin mention does not violate the Wigner–Eckart theorem, which prevents any reduced matrix element [the left-hand side (LHS) of the non-Abelian ETH] from depending on magnetic quantum numbers. Our $S_\th$ definition depends neither on the non-Abelian ETH's LHS nor on any magnetic spin quantum number.})
The density of states has the form
$d \coloneqq e^{ S_\th (\mathcal{E}, \mathcal{S}) } \, .$
The following section shows why we choose
these definitions: they enable the non-Abelian ETH to be self-consistent.

\section{Self-consistency of the non-Abelian ETH}
\label{sec_2s_main}

We argue analytically that the non-Abelian ETH is self-consistent in the sense of~\cite{Srednicki_96_Thermal}. There, Srednicki analyzed the square $A^2$ of a local operator $A$. The ETH implies directly and indirectly how matrix elements $(A^2)_{\alpha \alpha'}$ scale with system size. The two implications' scalings are the same, Srednicki showed. We review his argument in App.~\ref{sec_Consistency}. 
Here, we sketch the adaptation to SU(2) symmetry. Details appear in App.~\ref{app_2s}. The non-Abelian ETH is self-consistent, we show, if the thermodynamic entropy $S_\th$ is defined as in the previous subsection.

First, we introduce an analogue of $A^2$. Our argument requires an analogue because squaring one spherical tensor operator does not necessarily yield another such operator. Linearly combining such operators can, though~\cite{sakurai2017modern}: if 
$k \in \{|k_1-k_2|, |k_1-k_2| + 1, \ldots, k_1+k_2\}$,
\begin{align}
   \label{eq:combineAB}
   T_{q}^{(k)}
   =\sum_{q^{\prime}}\left\langle k, q \mid k_{1}, q^{\prime} ; k_{2}, q-q^{\prime}\right\rangle 
   A_{q^{\prime}}^{\left(k_{1}\right)} B_{q-q^{\prime}}^{\left(k_{2}\right)}\,.
\end{align}

The non-Abelian ETH directly implies through Eq.~\eqref{eq_NAETH} how $T_{q}^{(k)}$ scales. We focus on the block-diagonal reduced matrix element in this section:
\begin{align}
   \label{eq_NAETH_Direct_1}
   \langle \alpha || T^\KParen || \alpha \rangle
   & =  \mathcal{T}^\KParen ( \mathcal{E} , \mathcal{S} )
   \\ \nonumber & \quad \,
   +  e^{- S_\th ( \mathcal{E} , \mathcal{S} ) / 2 } \,
   f^{(T)}_{ \nu }  \left( \mathcal{E} , \mathcal{S} , \omega {=} 0  \right)  \,  R^{(T)}_{\alpha \alpha} \\
   \label{eq_NAETH_Direct_Scale}
   & = O(1)  +  O \big( \Dim^{-1/2} \big) \times O(1) \times O(1) \, .
\end{align}

Now, we must derive the non-Abelian ETH's indirect implication for how
$\langle \alpha || T^\KParen || \alpha \rangle$ scales.
This derivation is lengthier, so App.~\ref{app_2s} contains the details. We apply the non-Abelian ETH to the $A$ and $B$ in Eq.~\eqref{eq:combineAB}. Then, we invoke Clebsch–Gordan properties and assess how the resulting expression scales. This scaling must have the form in Eq.~\eqref{eq_NAETH_Direct_Scale}; such is the definition of an ETH statement's self-consistency~\cite{Srednicki_96_Thermal}.  
The non-Abelian ETH satisfies this requirement if $S_\th$ is defined as at the end of Sec.~\ref{sec_Backgrnd_NAETH}. The reason stems from a Clebsch–Gordan property that underlies selection rules in AMO physics: a Clebsch–Gordan coefficient
$\braket{s, m}{ s', m'; k, q}$ can be nonzero only if $m' = m - q$: of the $2 s' + 1$ possible $m'$ values, only one labels a nonzero coefficient. Appendix~\ref{app_2s} also extends the self-consistency argument to off--block-diagonal matrix elements.

\section{Model and methods}
\label{sec_Model}

Below, we describe the physical system we model, as well as its Hamiltonian and charges. Then, we describe our calculational strategy.

Consider a 1D chain of $N = 18$ qubits subject to open boundary conditions (the boundary spins do not interact with each other). Denote by $\sigma_a^{(j)}$ component $a \in \{x, y, z \}$ of qubit $j$. (We change notation from $S_a = \frac{1}{2} \sigma_a$ to Pauli operators for simplicity.) The system evolves under a Heisenberg model rendered nonintegrable by next-nearest-neighbor couplings:
\begin{align}
    \label{eq_Heis}
    H & = \sum_{j=1}^{N-1}  J_j  \,
    \vec{\sigma}^{(j)} \cdot \vec{\sigma}^{(j+1)} 
   + \sum_{j=1}^{N-2} J_j  \,
   \vec{\sigma}^{(j)} \cdot \vec{\sigma}^{(j+2)} \, .
\end{align}
The coupling strength $J_j = 1 + 0.3 \, \delta_{j3}$ is uniform except around site $j=3$. The offset precludes spatial-inversion symmetry, while the offset and boundary conditions preclude translational symmetry. These preclusions maintain our focus on non-Abelian symmetry. The offset's location and strength are arbitrary. Other choices yield qualitatively similar results when the offset is $>0.02$. 
Throughout this paper, identity operators $\id$ are implicitly tensored on wherever necessary so that operators are defined on the appropriate Hilbert spaces. 

We can now introduce the charges. Define component $a$ of the total spin as
$\sigma_a^\tot 
\coloneqq \sum_{j=1}^N  \sigma^{(j)}_a \, .$
The Hamiltonian conserves these components, which form the model's charges~\cite{NYH_20_Noncommuting,Kranzl_23_Experimental,NYH_22_How}:
$[ H ,  \sigma_a^\tot ] = 0$ $\, \forall a = x, y, z$.
We sometimes refer to the single-qubit operators $\sigma_a$ as charges for convenience, although $H$ conserves only the global $\sigma_a^\tot$s.
The charges do not commute with each other:
$[\sigma_a , \sigma_b] \neq 0$ if $a \neq b$.

We perform our calculations by exact diagonalization. 
We parallelize and optimize large-sparse-matrix operations using the Intel Math Kernel Library.\footnote{
We identify certain equipment or software to specify the numerical procedure adequately. Such identification does not imply a recommendation or endorsement of any product or service by NIST; nor does it imply that the equipment or software is necessarily the best available for the purpose.}
We calculate as follows the eigenbasis $\{ \ket{\alpha, m} \}$ shared by $H$ , $\vec{S}^2$, and $S_z$.
First, we calculate an eigenbasis shared by $\vec{S}^2$ and $S_z$ but not $H$: we construct each basis element by forming a linear combination of $N$-qubit tensor products, using Clebsch–Gordan coefficients as expansion coefficients. Denote this basis by $\{ \ket{s_\alpha, m, n} \}$; the $n$ resolves the degeneracies. Second, we calculate the matrix elements $\bra{s_\alpha, m, n} H \ket{s_\alpha, m, n'}$. Equivalently, we represent $H$ as a matrix relative to the initial eigenbasis but ignore the off--block-diagonal elements. One $(s_\alpha, m)$ value labels each diagonal block. We diagonalize each block, calculating the joint eigenstates $\ket{\alpha, m}$.
This block-centric strategy saves computational resources.

In terms of $\{ \ket{\alpha, m} \}$, we represent operators $T^\KParen_q$ as matrices, to calculate reduced matrix elements. Throughout the next section, we analyze the representative example
$T^\1_0 = \frac{1}{2} \sigma^{( \lceil N/2 \rceil )}_z$.
We choose $q=0$ for computational convenience (App.~\ref{app:data}). However, App.~\ref{app:data} exhibits data about other $T^\KParen_q$s.

Appendix~\ref{app_JDN_Num} complements the main text's model with a 1D Heisenberg model subject to periodic boundary conditions, as well as translational and spatial-inversion symmetries. The extra symmetries enable calculations about a much larger system: $N = 26 > 18$. The results qualitatively resemble the following section's results. While buttressing our main findings, the appendix suggests the non-Abelian ETH's generality.

\section{Numerical support for the non-Abelian ETH}
\label{sec_results}

Our main results numerically evidence that the non-Abelian ETH models local operators' reduced matrix elements. We begin with a two-part prelude (Sec.~\ref{sec_Prelude}): first, we confirm that $H$ is nonintegrable and so should obey the non-Abelian ETH. Second, we present the DOS within each $\vec{S}^2$ eigenspace.

After the prelude, we analyze block-diagonal reduced matrix elements of $T^\1_0$ in two ways (Sec.~\ref{sec_Block}). First, we show that the elements follow a Gaussian distribution, consistently with the non-Abelian ETH's $R^{(T)}_{\alpha \alpha'} \, .$ Second, we present data consistent with smooth dependence of $\mathcal{T}^\1$ on $\mathcal{E}$ and $\mathcal{S}$ in the thermodynamic limit. We analyze off--block-diagonal reduced matrix elements and $f^{(T)}_\nu$ similarly (Sec.~\ref{sec_Off_Block}).

We complement these qualitative results with three quantitative tests of the non-Abelian ETH (Sec.~\ref{sec_Quant}). All the tests involve the variances of matrix elements (i) drawn from narrow energy windows and (ii) associated with fixed $s_\alpha$ values. First, we show that the ratio (variance of block-diagonal elements)/(variance of off--block-diagonal elements), averaged appropriately, equals 2~\cite{DAlessio_16_From}. Second, we demonstrate that $T^\1_0$'s block-diagonal reduced matrix elements have a variance that decays inversely with the DOS. Third, we show that the off--block-diagonal reduced matrix elements' variance decays similarly.

Appendix~\ref{app:data} presents data about operators $T^\KParen_q$ other than $T^\1_0$, as mentioned in the previous section. Some of the operators have different $q$ values. To within numerical precision, we confirm, $f^{(T)}_\nu$ does not depend on $q$, as the Wigner--Eckart theorem requires of the non-Abelian ETH.

\subsection{Prelude to numerical support}
\label{sec_Prelude}

Before testing the non-Abelian ETH, we perform two warm-up calculations. First, we show that $H$ is nonintegrable---that one should expect $H$ to obey the non-Abelian ETH. Then, we calculate the DOS, used in future calculations.

To demonstrate our model's nonintegrability, we calculate energy-gap statistics~\cite{DAlessio_16_From}: within each joint eigenspace shared by $\vec{S}^2$ and $S_z$, we calculate the gaps between consecutive eigenenergies.\footnote{
The Hamiltonian's symmetry complicates the statistics of gaps calculated from multiple eigenspaces' energies~\cite{Giraud_22_GapRatio,JaeDongXXZ}.}

\begin{figure}[ht]
    \includegraphics[width=0.95\linewidth]{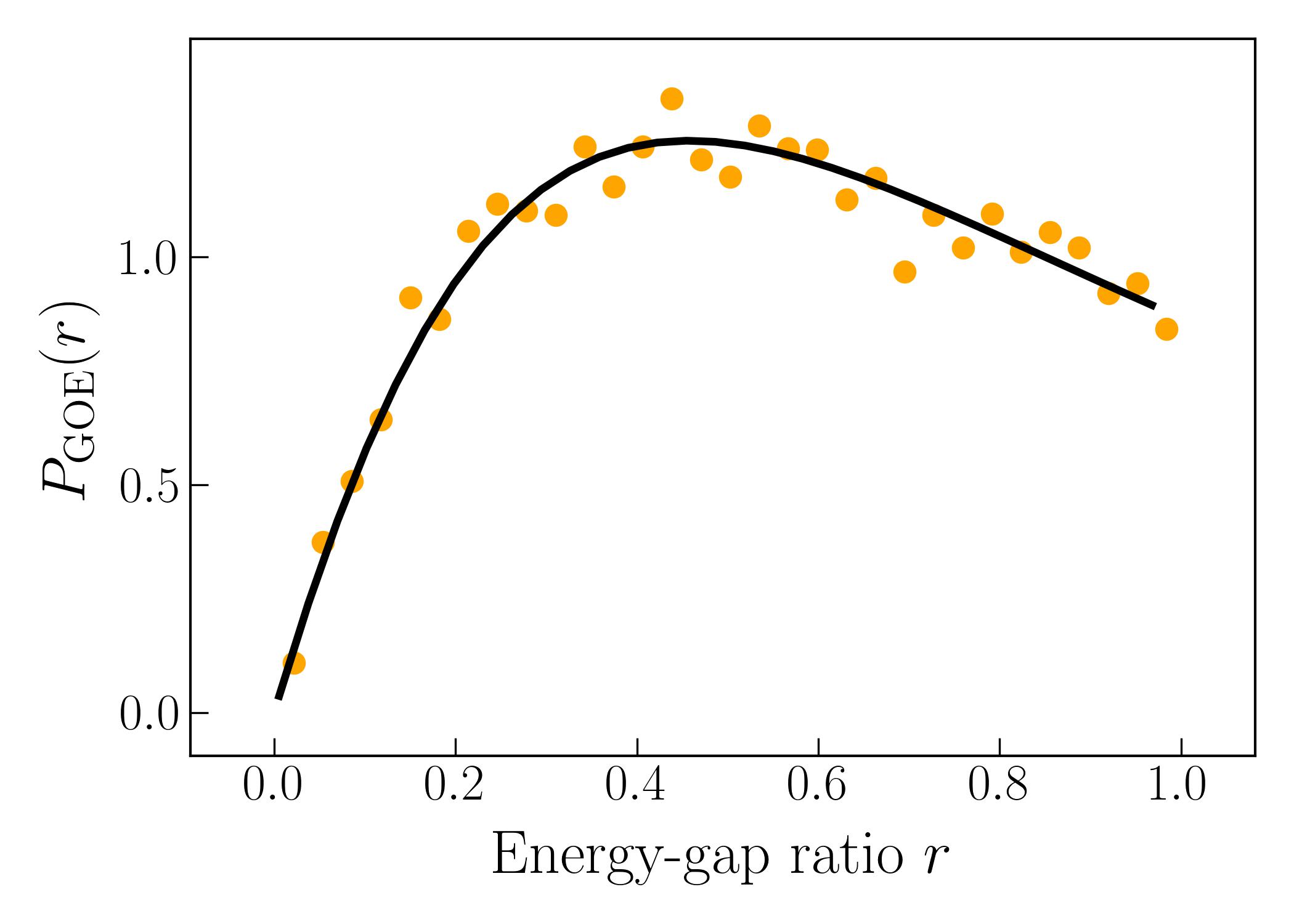}
    \caption{Probability density of the minimal gap ratio $r$ in an 18-qubit system's ($s_\alpha {=}3, m{=}0$) eigenspace. The random-matrix-theory prediction $P_{\rm GOE} (r)$ fits the histogram well, indicating the Hamiltonian's nonintegrability. The GOE fit's coefficient of determination is $R^2=0.921$.
    }
    \label{fig:Q18_Energy_gaps_s3m0}
\end{figure}

From each gap pair $(E_{n+1} - E_n, \, E_n - E_{n-1})$, we compute the minimal gap ratio~\cite{Oganesyan_07_Localization}
$\min\left\{ \frac{E_{n+1}-E_n}{E_n-E_{n-1}}, \frac{E_n-E_{n-1}}{E_{n+1}-E_n} \right\}$.
We then calculate the probability distribution that any given minimal gap ratio (in the subspace) is of size $r$. 
Figure~\ref{fig:Q18_Energy_gaps_s3m0} illustrates the gap-ratio probability with 
the ($s_\alpha {=}3, m{=}0)$ eigenspace. The black curve represents the probability density predicted via the Gaussian orthogonal ensemble: 
$P_{\rm GOE} (r) = \frac{27}{4}\frac{r(1+r)}{(1+r+r^2)^{5/2}}$~\cite{Atas.2013,DAlessio_16_From}.
This prediction suits our model because the Heisenberg Hamiltonian~\eqref{eq_Heis} has time-reversal symmetry and is represented, relative to the $\sigma_z$-product basis, by a real, symmetric matrix.
The distribution fits the histogram well, evidencing nonintegrability; the linear-regression coefficient of determination is $R^2 = 0.921$ . Similar results follow from other joint eigenspaces that contain at least a few hundred joint eigenstates each. Reference~\cite{Protopopov_20_Non} contains similar results.

\begin{figure}[ht]
    \centering
    \includegraphics[width=0.95\linewidth]{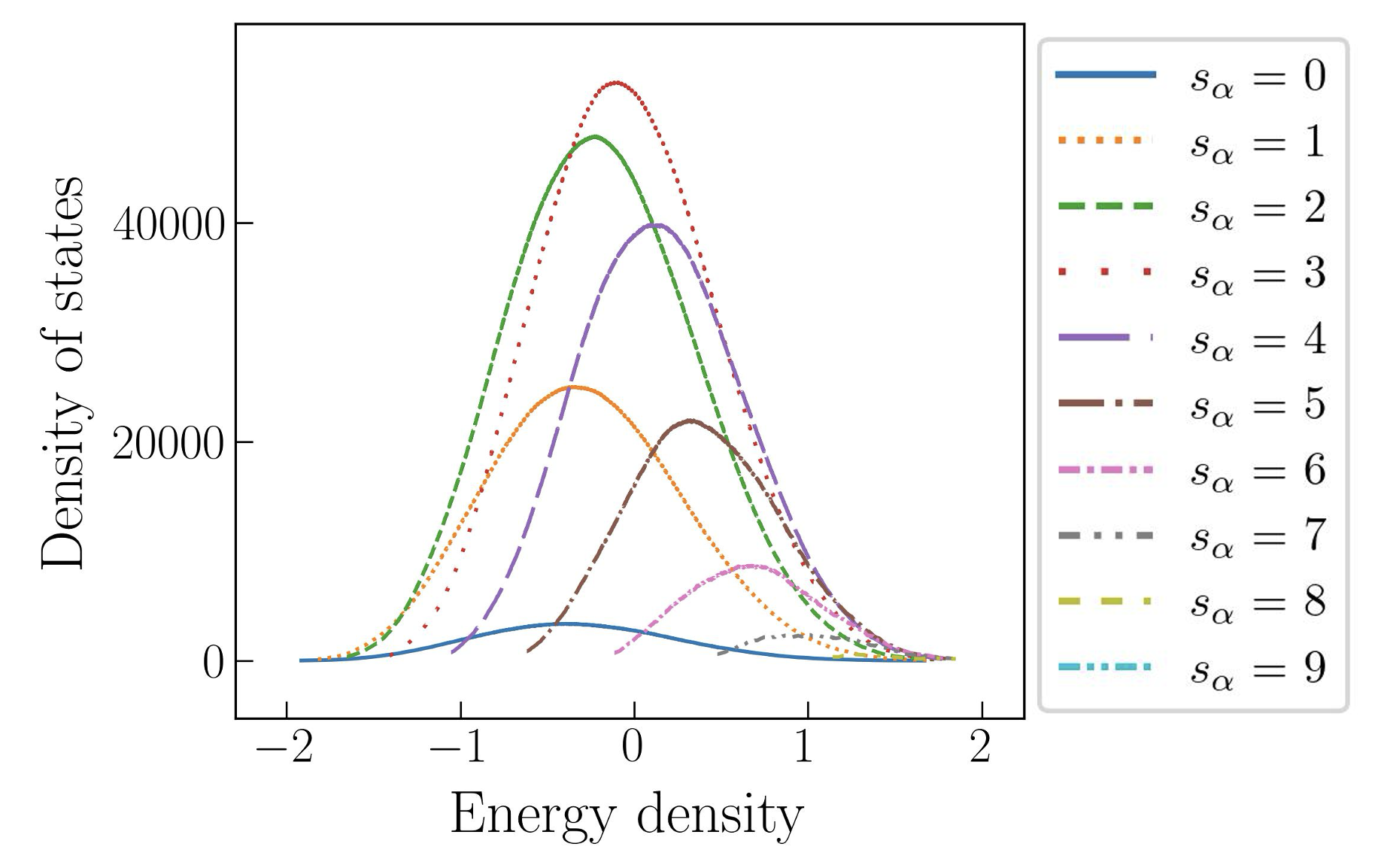}
    \caption{Density of states of the $N{=}18$
    Heisenberg chain. Different curves correspond to different $\vec{S}^2$ eigenspaces. 
    }
    \label{fig:Q18_DOS}
\end{figure}

Having demonstrated our model's nonintegrability, we calculate its DOS (Fig.~\ref{fig:Q18_DOS}). The non-Abelian ETH does not predict about the DOS. However, we use these data to identify the energy $E_\alpha$, within each $\vec{S}^2$ eigenspace, where the DOS peaks. We also use the DOS to analyze reduced matrix elements' scalings.

\subsection{Block-diagonal reduced matrix elements}
\label{sec_Block}

\begin{figure}[ht]
    \includegraphics[width=0.95\linewidth]{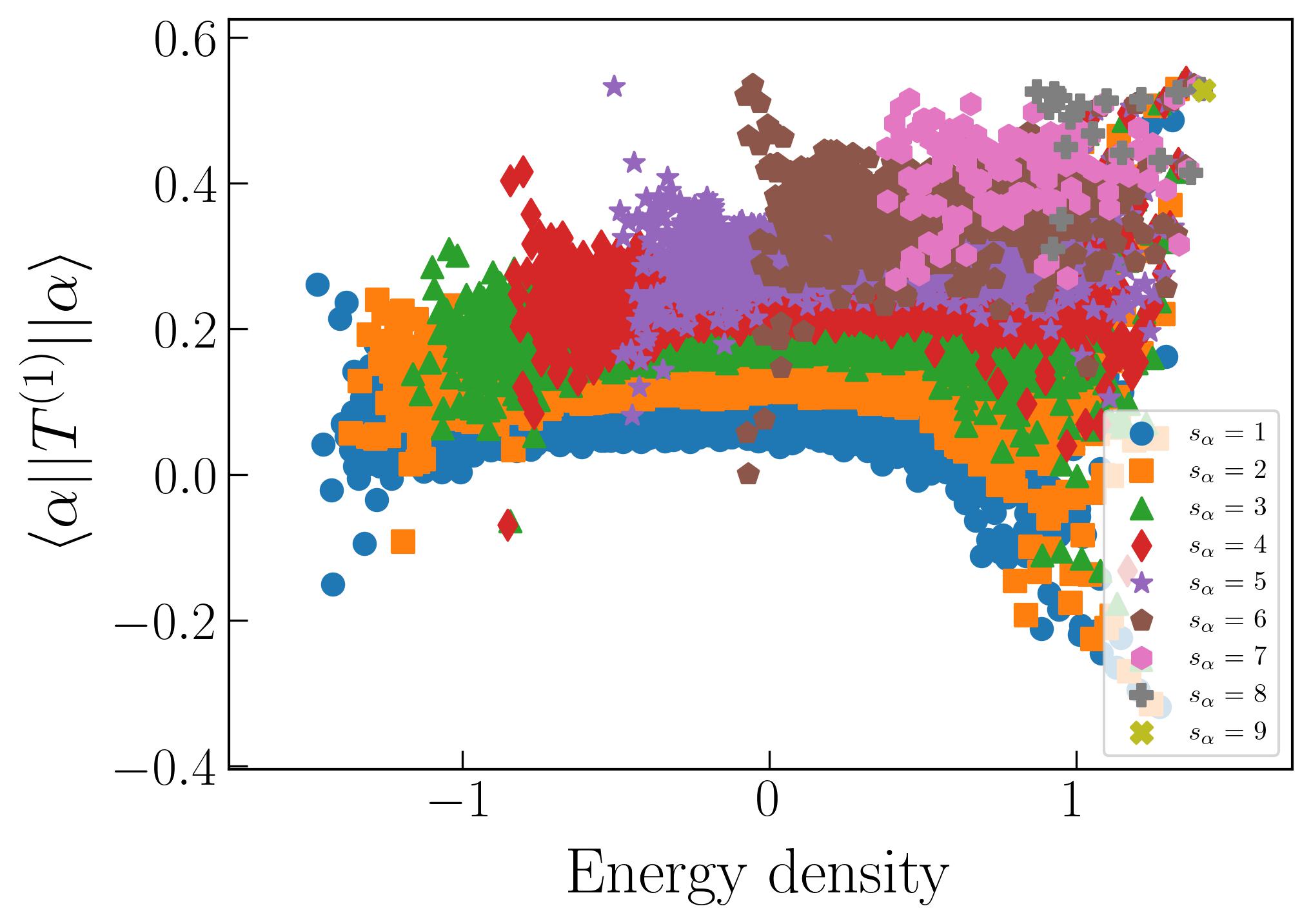}
    \caption{Block-diagonal reduced matrix elements 
    $\langle \alpha || T^\1 || \alpha \rangle$ versus energy density, for $N=18$. The elements form bands labeled by $s_\alpha$. Furthermore, 
    $\langle \alpha || T^\1 || \alpha \rangle$ varies with $s_\alpha/N$ in a manner suggestive of smoothness.}
    \label{fig:Q18_Energy_gaps_GEO}
\end{figure}

We now exhibit properties of the block-diagonal reduced matrix elements $\langle \alpha || T^\1 || \alpha \rangle$. Figure~\ref{fig:Q18_Energy_gaps_GEO} shows these elements plotted against energy density. 
The markers form bands distinguished by $s_\alpha$, represented by color (and marker shape). Within each band, $\langle \alpha || T^\1 || \alpha \rangle$ varies in a manner suggestive of smoothness (as a function of the energy density $E_\alpha/N$) in the thermodynamic limit. 
Each band has a width, or nonzero variance, about the mean $\langle \alpha || T^\1 || \alpha \rangle$ value. Furthermore, at any given energy density $E_\alpha/N$, 
$\langle \alpha || T^\1 || \alpha \rangle$ varies with $s_\alpha/N$ in a manner indicative of smoothness. 

Figure~\ref{fig:DiagShaded} in App.~\ref{app:data} further supports the expectation of smoothness. The figure shows the reduced matrix elements' averages over narrow energy-density windows. At each $s_\alpha$ value, the mean varies mostly linearly with $E_\alpha / N$, fluctuating only slightly. Furthermore, as a function of $s_\alpha$, the mean varies monotonically and nearly linearly. These behaviors are consistent with the non-Abelian ETH's first term (Sec.~\ref{sec_Backgrnd_NAETH}).

\begin{figure}[ht]
    \centering
    \includegraphics[width=0.95\linewidth]{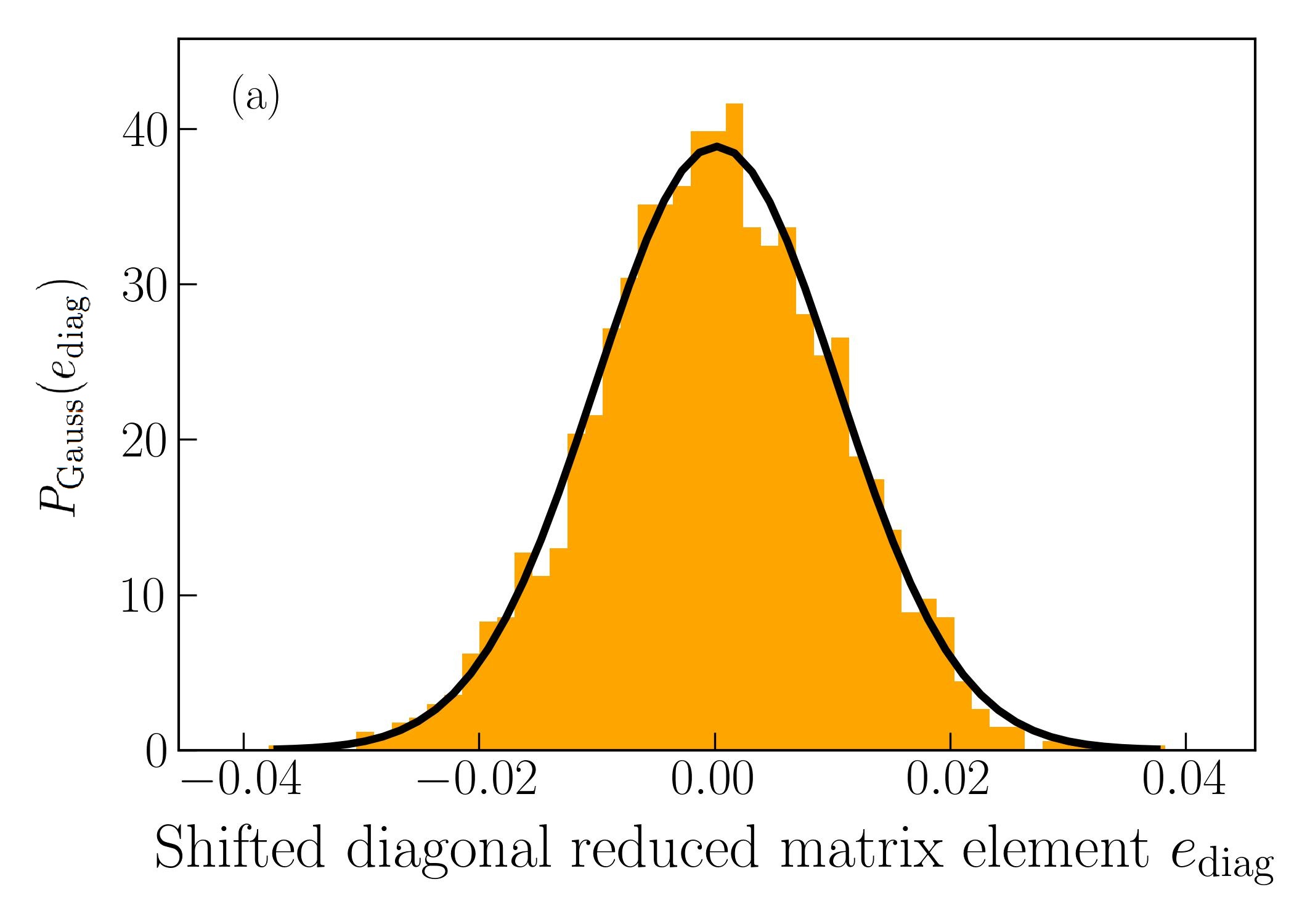}
    \caption{Histogram of block-diagonal reduced matrix elements, shifted so that the mean element vanishes. 
    The shifted matrix elements are expected to equal the non-Abelian ETH's second term. 
    The eigenenergies $E_\alpha$ come from a narrow energy-density window (of width 0.1).
    The energy eigenstates used come from 
    the representative $s_\alpha {=}3$ band (Fig.~\ref{fig:Q18_Energy_gaps_GEO}). 
    A Gaussian distribution $P_\mathrm{G}$ fits the histogram,
    as expected from the non-Abelian ETH's $R^{(T)}_{\alpha \alpha} \, .$ The Gaussian distribution's fit has a coefficient of determination $R^2=0.989$.}
    \label{fig:Q18_Diag_Hist_s3}
\end{figure}

We now present evidence for the non-Abelian ETH's second term [Eq.~\eqref{eq_NAETH}], focusing on block-diagonal reduced matrix elements $\langle \alpha || T^\1 || \alpha \rangle$. We focus on one $\vec{S}^2$ eigenspace and on a small energy-density window. The window has a width of 0.1.\footnote{
For reference, the  Hamiltonian has a bandwidth of 68. The window-width 0.1 balances the need for a narrow window with the need for enough reduced matrix elements that their variance is meaningful. To secure enough matrix elements, we center the window near the peak of the eigenspace's DOS (Fig.~\ref{fig:Q18_DOS}).}
We subtract the average reduced matrix element, expected to equal
$\mathcal{T}^\1( \mathcal{E}, s_\alpha )$, from 
$\langle \alpha || T^\1 || \alpha \rangle$. The difference should equal the non-Abelian ETH's second term. Due to the 
$R^{(T)}_{\alpha \alpha}$ in Eq.~\eqref{eq_NAETH}, the difference should behave as a random variable. Figure~\ref{fig:Q18_Diag_Hist_s3} supports this expectation with a histogram from the representative $s_\alpha {=} 3$ subspace. $e_\mathrm{diag}$ denotes a possible value of any given block-diagonal element, minus the average block-diagonal element.  A Gaussian distribution $P_\mathrm{Gauss}(e_\mathrm{diag})$ fits the histogram well; the coefficient of determination is $R^2 = 0.989$.

\subsection{Off--block-diagonal reduced matrix elements}
\label{sec_Off_Block}

\begin{figure}[ht]
    \centering
    \includegraphics[width=0.95\linewidth]{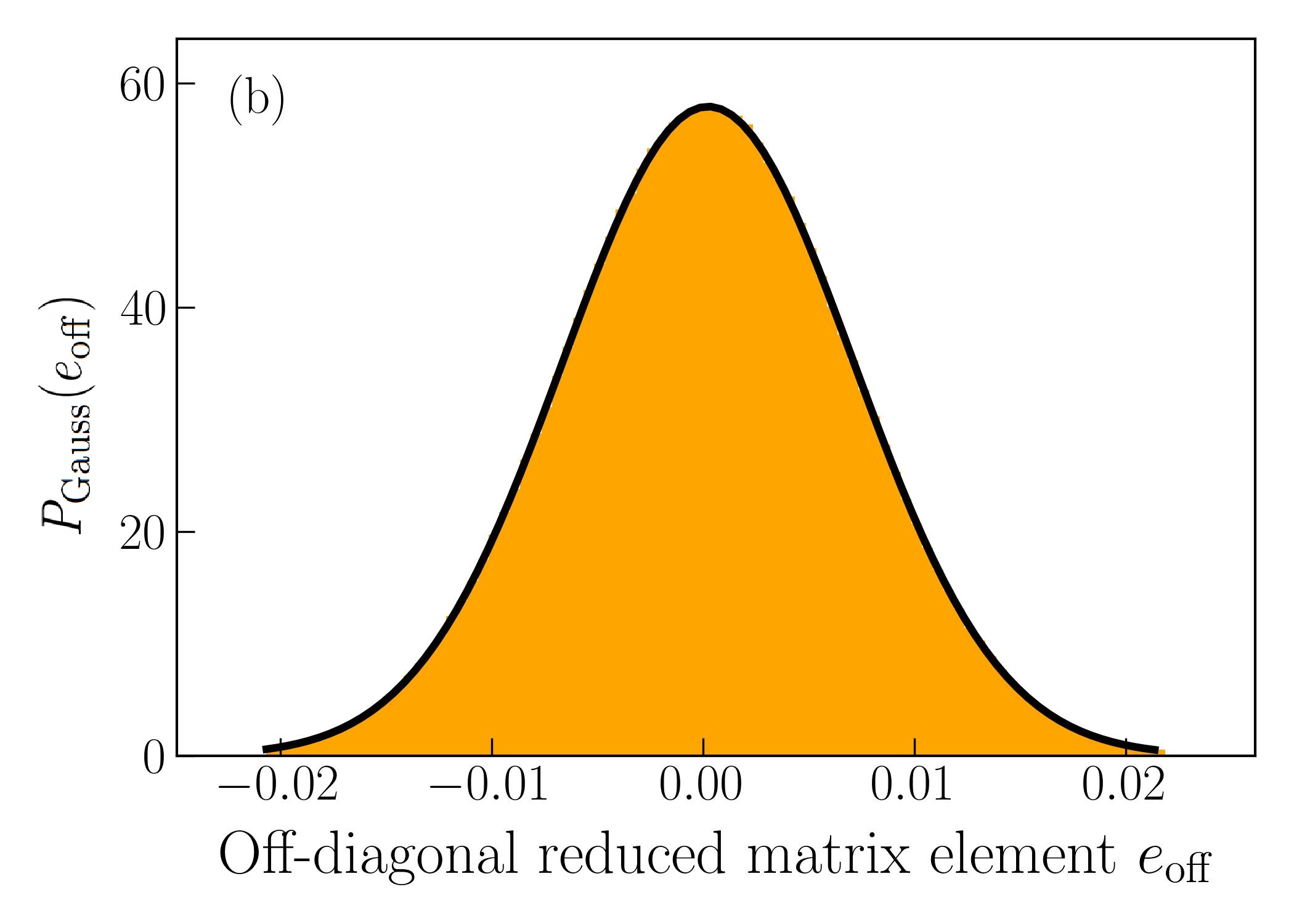}
    \caption{Histogram of off--block-diagonal reduced matrix elements $\langle \alpha || T^\1 || \alpha' \rangle$. We collected the statistics as described under Fig.~\ref{fig:Q18_Diag_Hist_s3}. A Gaussian distribution fits the histogram, supporting the non-Abelian ETH's second term. The Gaussian distribution's fit has a coefficient of determination $R^2=0.999$.
    }
    \label{fig:Q18_Offdiag_Hist_s3}
\end{figure}

Having analyzed block-diagonal reduced matrix elements, we exhibit properties of off--block-diagonal elements. These elements should vary randomly, due to the $R^{(T)}_{\alpha \alpha'}$ in the non-Abelian ETH [Eq.~\eqref{eq_NAETH}]. Figure~\ref{fig:Q18_Offdiag_Hist_s3} supports this prediction with a histogram of elements
$\langle \alpha || T^\1 || \alpha' \rangle$. The elements come from a narrow energy-density window within the $s_\alpha{=}3$ subspace. The window is centered at $\omega=0$ and at the $\mathcal{E}/N$ value where the DOS maximizes within the subspace. The window has a width of $0.1$ in $E_\alpha/N$ and in $E_{\alpha'}/N$.
A Gaussian distribution $P_\mathrm{Gauss}(e_\mathrm{off})$ models the histogram well, consistently with the non-Abelian ETH's $R^{(T)}_{\alpha \alpha'} \, .$  

\begin{figure}[ht]
    \centering
    \includegraphics[width=0.95\linewidth]{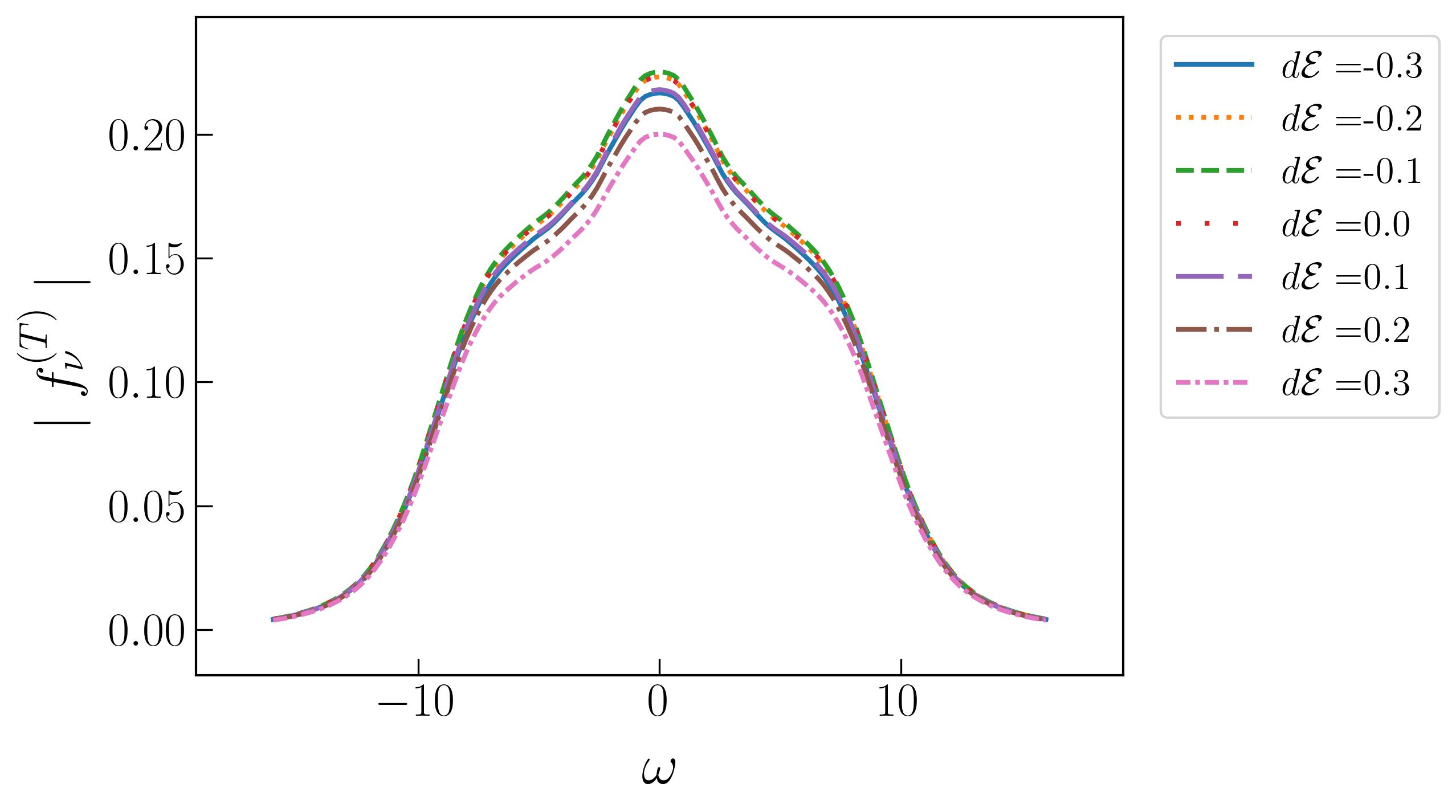}
    \caption{Magnitude of the function 
    $f^{(T)}_{\nu=0}\left( \mathcal{E}, \mathcal{S}{=}3, \omega \right)$, for the operator $T^{(1)}_0$, as a function of $\omega$. Different curves correspond to different offsets $d\mathcal{E}$ of $\mathcal{E}$ from the energy that maximizes the DOS at a given $\mathcal{S}$. These 18-qubit data are consistent with the function's being smooth in both arguments in the thermodynamic limit.}
    \label{fig:ffunctE_T10}
\end{figure}

\begin{figure}[ht]
    \centering
    \includegraphics[width=0.95\linewidth]{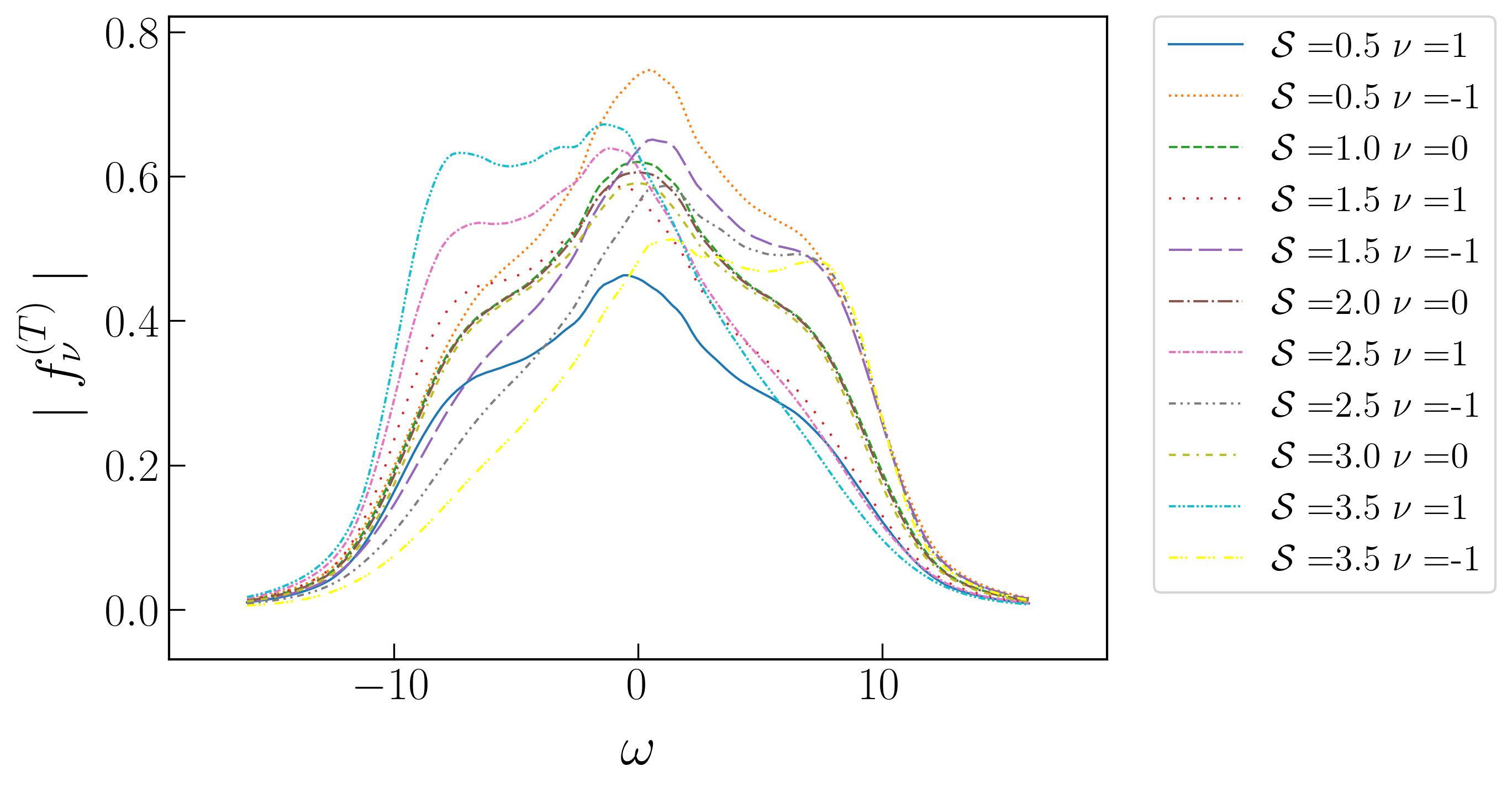}
    \caption{Magnitude of $f^{(T)}_\nu \left( \mathcal{E} {=} 0, \mathcal{S}, \omega \right)$, for the operator $T^{(1)}_0$, as a function of $\omega$. Different curves correspond to different $\mathcal{S}$ and $\nu$ values. If $\nu=0$, then $| f^{(T)}_\nu |$ is symmetric with respect to reflection about $\omega=0$. These 18-qubit numerics suggest that $| f^{(T)}_\nu |$ may be a smooth function of $\omega$ and $\mathcal{S}$ in the thermodynamic limit. }
    \label{fig:ffunctS_T10}
\end{figure}

The off--block-diagonal elements depend not only on the random $R^{(T)}_{\alpha \alpha'}$, but also on the smooth 
$f^{(T)}_{ \nu }  (\mathcal{E}, \mathcal{S}, \omega)$. We calculate the latter's square modulus as follows. First, we identify a narrow energy-density window about $\mathcal{E}/N$. The window has a width of 0.1 in $E_\alpha/N$ and in $E_{\alpha'}/N$. Within the window are the eigenenergy densities associated with the Hamiltonian eigenstates ($\ket{\alpha, m}$ and $\ket{\alpha', m'}$) that define the off--block-diagonal reduced matrix elements. We calculate these elements' variance. Then, we divide by the DOS at $\mathcal{E}$.\footnote{
The DOS is defined at the end of Sec.~\ref{sec_Backgrnd_NAETH}, in terms of the magnetic quantum number 0.}$^,$\footnote{
The non-Abelian ETH~\eqref{eq_NAETH} reveals why we divide by the DOS: the reduced matrix elements' root-mean-square differs from $f_\nu^{(T)}(\mathcal{E}, \mathcal{S}, \omega)$ by $e^{-S_\th( \mathcal{E}, \, \mathcal{S}) / 2}$---by a function of the DOS.}
Figure~\ref{fig:ffunctE_T10} highlights the function's dependence on $\mathcal{E}$ and $\omega$; and Fig.~\ref{fig:ffunctS_T10}, the dependence on $\mathcal{S}$ and $\omega$. Such finite-size-system numerics cannot exhibit smooth dependences. Evidencing smooth dependence on $\mathcal{S}$ is most difficult: consecutive $s_\alpha$ values differ by 1, while consecutive $E_\alpha$ values differ by $O(N / 2^N) \ll 1$.\footnote{ 
The bandwidth is $O(N)$, and the number of eigenenergies is $2^N$.}
However, the figures suggest that $f^{(T)}_\nu \left( \mathcal{E}, \mathcal{S}, \omega \right)$ may vary smoothly with its arguments in the thermodynamic limit. It satisfies objective measures of smoothness: the first and second derivatives with respect to $\omega$ and $\mathcal{E}$ are well-behaved. In addition, a low-order polynomial fits the data well, with low residuals. (The polynomial is in $\omega$, if $\mathcal{E}$ is fixed, or in $\mathcal{E}$, if $\omega$ is fixed.) Smoothness in $\mathcal{S}$ is especially difficult to test far from the thermodynamic limit: $\mathcal{S}$ is discrete, so significant gaps separate the curves in Fig.~\ref{fig:ffunctS_T10}. However, the curves' shapes change little as $\mathcal{S}$ varies: each curve is a single- or double-peaked function of $\omega$. Each curve decays rapidly as $| \omega |$ grows beyond 10. Furthermore, the peaks' $y$-coordinates lie within a width-0.3 window.

$f^{(T)}_\nu$ even exhibits a subtlety discussed in~\cite{DAlessio_16_From,Srednicki_99_Approach,Khatami_13_Fluctuation}: 
$f^{(T)}_\nu$ is $O(1)$ only near $\omega = 0$. As $\omega$ grows beyond $O(1)$, $f^{(T)}_\nu$ decays quickly. The reason is, a local operator $T^\1_0$ has a high probability of effecting transitions only between $\ket{\alpha m}$s associated with nearby $E_\alpha$s (and so small $\omega$s).

\subsection{Quantitative tests of the non-Abelian ETH}
\label{sec_Quant}

We have shown that block-diagonal reduced matrix elements have qualitative properties predicted by the non-Abelian ETH, as do off--block-diagonal elements. We now combine the two element types in three quantitative tests, all involving reduced matrix elements' variances.

\begin{figure}[ht]
    \centering
    \includegraphics[width=0.95\linewidth]{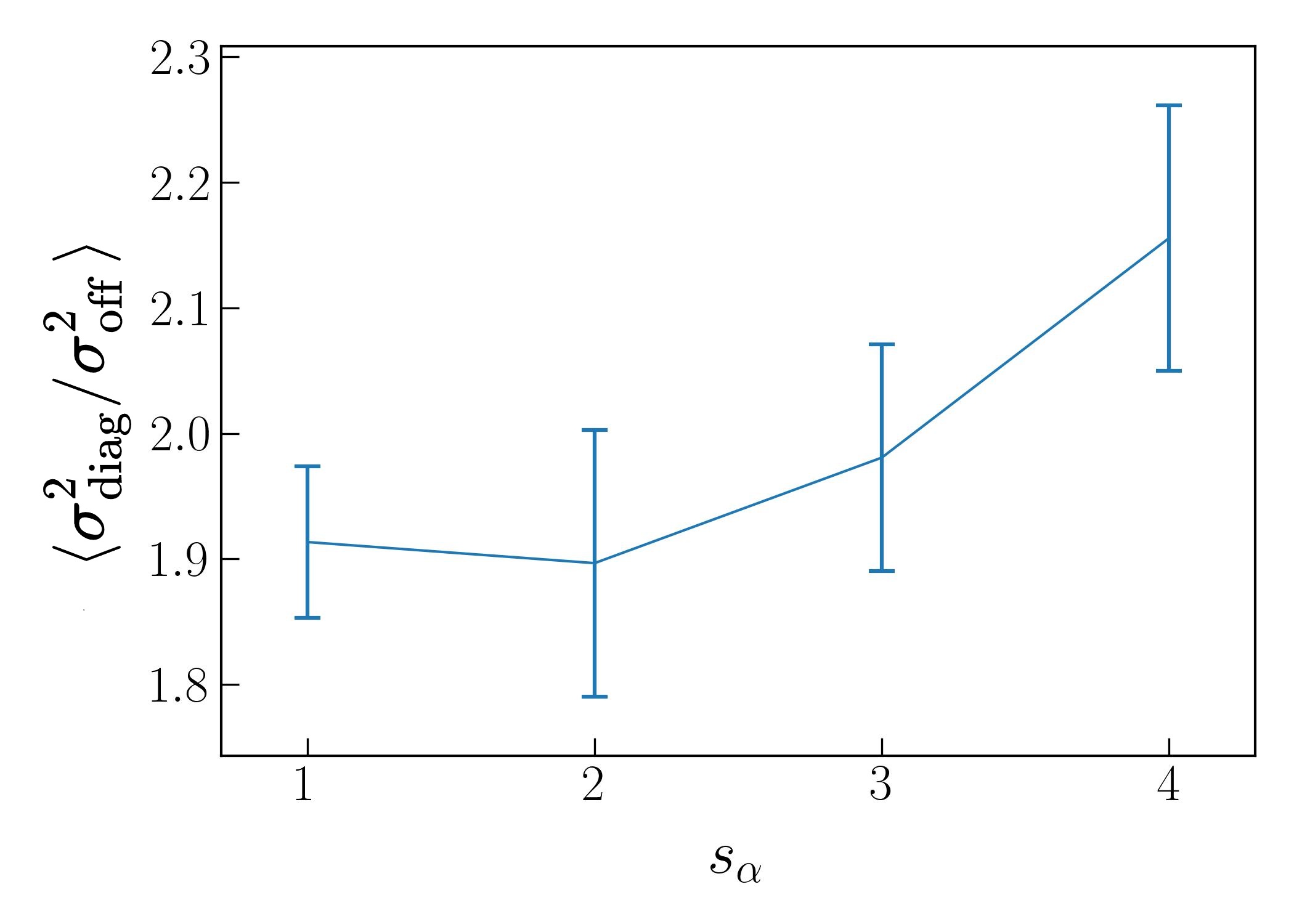}
    \caption{Average of $\sigma_\text{diag}^2 / \sigma_\text{off}^2$, as a function of the spin quantum number $s_\alpha$, for the operator $T_0^{(1)}$. The average is over many narrow energy-density windows. 
    For each $s_\alpha$, the average ratio lies close to the value predicted with random-matrix theory, 2. Furthermore, averaging $\sigma_\text{diag}^2 / \sigma_\text{off}^2$ over $s_\alpha$ yields $1.99 \pm 0.10$.}
    \label{fig:varratios_T10}
\end{figure}

The first test features a ratio of variances. Within each $\vec{S}^2$ eigenspace, we calculate the reduced matrix elements
$\langle \alpha || T^\1 || \alpha' \rangle$ defined in terms of eigenstates ($\ket{\alpha, m}$ and $\ket{\alpha', m'}$) associated with energy densities ($E_\alpha/N$ and $E_{\alpha'}/N$) within a small window. Whenever we write \emph{elements} within this subsection, we mean \emph{reduced matrix elements}. We calculate the block-diagonal elements' variance, $\sigma^2_\text{diag}$, and the off--block-diagonal elements', $\sigma^2_\text{off}$.  
Then, we repeat this process with nearby energy-density windows.\footnote{
All the narrow windows come from an encompassing window of width (in energy-density space) 0.5. We choose 0.5 because the DOS does not decrease significantly within a width-0.5 window centered on the DOS's peak. Figure~\ref{fig:Q18_DOS} evidences this behavior, exemplified by the $s_\alpha{=}3$ curve.}
We average $\sigma^2_\text{diag} / \sigma^2_\text{off}$ across the windows. The averaged ratio should equal 2, by random-matrix theory~\cite{JaeDongXXZ,DAlessio_16_From}.
Figure~\ref{fig:varratios_T10} shows the average as a function of $s_\alpha$. The ratio within each $s_\alpha$ subspace is sensitive to the subspace's dimensionality. However, the average of $\sigma_\text{diag}^2 / \sigma_\text{off}^2$ over $s_\alpha$ values equals 2, to within the standard deviation: $1.99 \pm 0.10$.
The data match the prediction best when taken from an $s_\alpha$ eigenspace that contains enough eigenstates. We therefore use the $s_\alpha \in \{1, 2, \ldots, 4\}$ eigenspaces.

\begin{figure}[ht]
    \centering
    \includegraphics[width=0.95\linewidth]{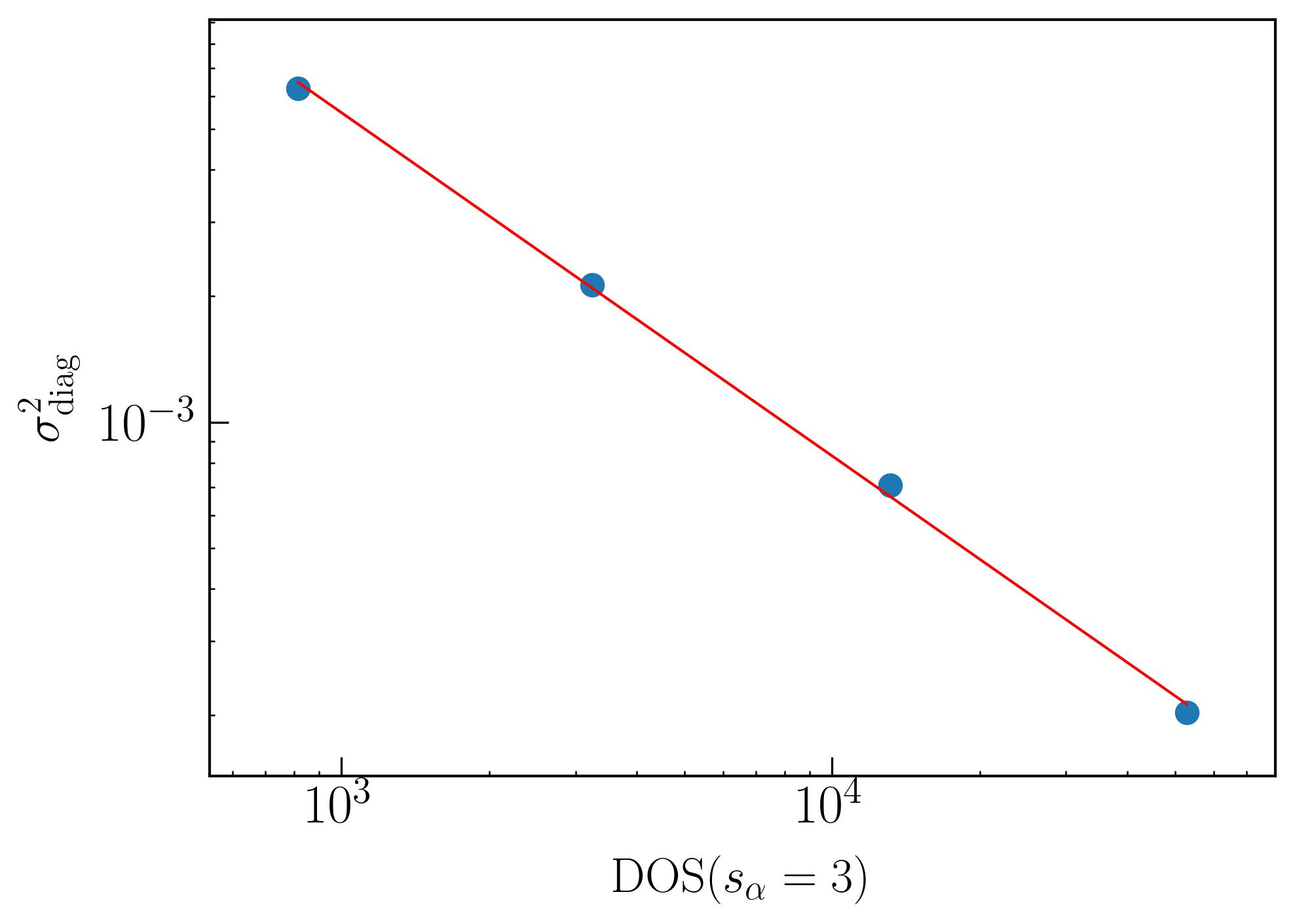}
    \caption{Block-diagonal reduced matrix elements' variance, $\sigma_{\rm diag}^2$, versus DOS. We sample the elements at $s_\alpha=3$ and at the $\mathcal{E}$ value that maximizes the DOS.  The data describe $T^{(1)}_0$ and system sizes $N=12$–18. The log–log plot displays a linear relationship, which holds in the other parameter regimes checked. The slope is approximately $-0.83 \pm 0.02$, as compared with the prediction $-1$ from the non-Abelian ETH.
    }
    \label{fig:ExpDecayDiag}
\end{figure}

Figure~\ref{fig:ExpDecayDiag} displays our second quantitative test of the non-Abelian ETH. The figure shows the reduced block-diagonal matrix elements' variance, $\sigma_{\rm diag}^2$, within a small energy-density window (of width 0.1). We sample the elements at $s_\alpha=3$, near the $\mathcal{E}/N$ value that maximizes the DOS. This analysis relies on the DOS data in Fig.~\ref{fig:Q18_DOS}.

Figure~\ref{fig:ExpDecayDiag} shows $\sigma_{\rm diag}^2$ as a function of the DOS on a log--log plot.
The variance decays algebraically in the DOS, with a slope of $-0.83 \pm 0.02$.  The non-Abelian ETH [Eq.~\eqref{eq_NAETH}] predicts a slope of $-1$. We justify this claim with an argument that governs elements selected from a narrow energy window at a fixed $s_\alpha {=} s_{\alpha'}$ value, regardless of whether the elements are on or off the block-diagonal. The reduced matrix element has a variance
\begin{align}
   \label{eq_var_help1}
   & {\rm var} \left( \langle \alpha || T^\KParen || \alpha' \rangle \right) 
   \\ \nonumber
   & = \mathbb{E} \left( \left[
   \langle \alpha || T^\KParen || \alpha' \rangle
   - \mathbb{E} \left( \langle \alpha || T^\KParen || \alpha' \rangle \right)
   \right]^2  \right) .
\end{align}
The average $\mathbb{E}$ is over the energy window. Let us substitute in from the non-Abelian ETH~\eqref{eq_NAETH}. Throughout the energy window, $\mathcal{E}$ and $\omega$ are approximately constant. By design, $\mathcal{S} = s_\alpha = s_{\alpha'}$ is constant. Therefore, the smooth functions $\mathcal{T}^\KParen (\mathcal{E}, \mathcal{S}, \omega)$, $S_\th (\mathcal{E}, \mathcal{S})$,
and $f_\nu^{(T)} (\mathcal{E}, \mathcal{S}, \omega)$ are approximately constant. In contrast, $R_{\alpha, \alpha'}^{(T)}$ varies considerably. Equation~\eqref{eq_var_help1} therefore simplifies to
\begin{align}
   & {\rm var} \left( \langle \alpha || T^\KParen || \alpha' \rangle \right)
   \\ \nonumber &
   \label{eq_var_help2}
   \approx e^{- S_\th (\mathcal{E}, \mathcal{S}) }
   \left[ f_\nu^{(T)} (\mathcal{E}, \mathcal{S}, \omega) \right]^2
   {\rm var} \left( R_{\alpha \alpha'}^{(T)} \right) .
\end{align}
The final factor equals 1 (Sec.~\ref{sec_Backgrnd_NAETH}). 
The first factor equals $d^{-1}$, by the final paragraph of Sec.~\ref{sec_Backgrnd_NAETH}. Let us take the log of each side of Eq.~\eqref{eq_var_help2}:
\begin{align}
   \log \left( 
   {\rm var} \left( 
   \langle \alpha || T^\KParen || \alpha' \rangle 
   \right) \right)
   \approx - \log(d) + \const
\end{align}
Hence the non-Abelian ETH predicts a slope of $-1$ for Fig.~\ref{fig:ExpDecayDiag}. Finite-size effects appear to prevent the calculated slope from equaling -1 precisely.

\begin{figure}[ht]
    \centering
    \includegraphics[width=0.95\linewidth]{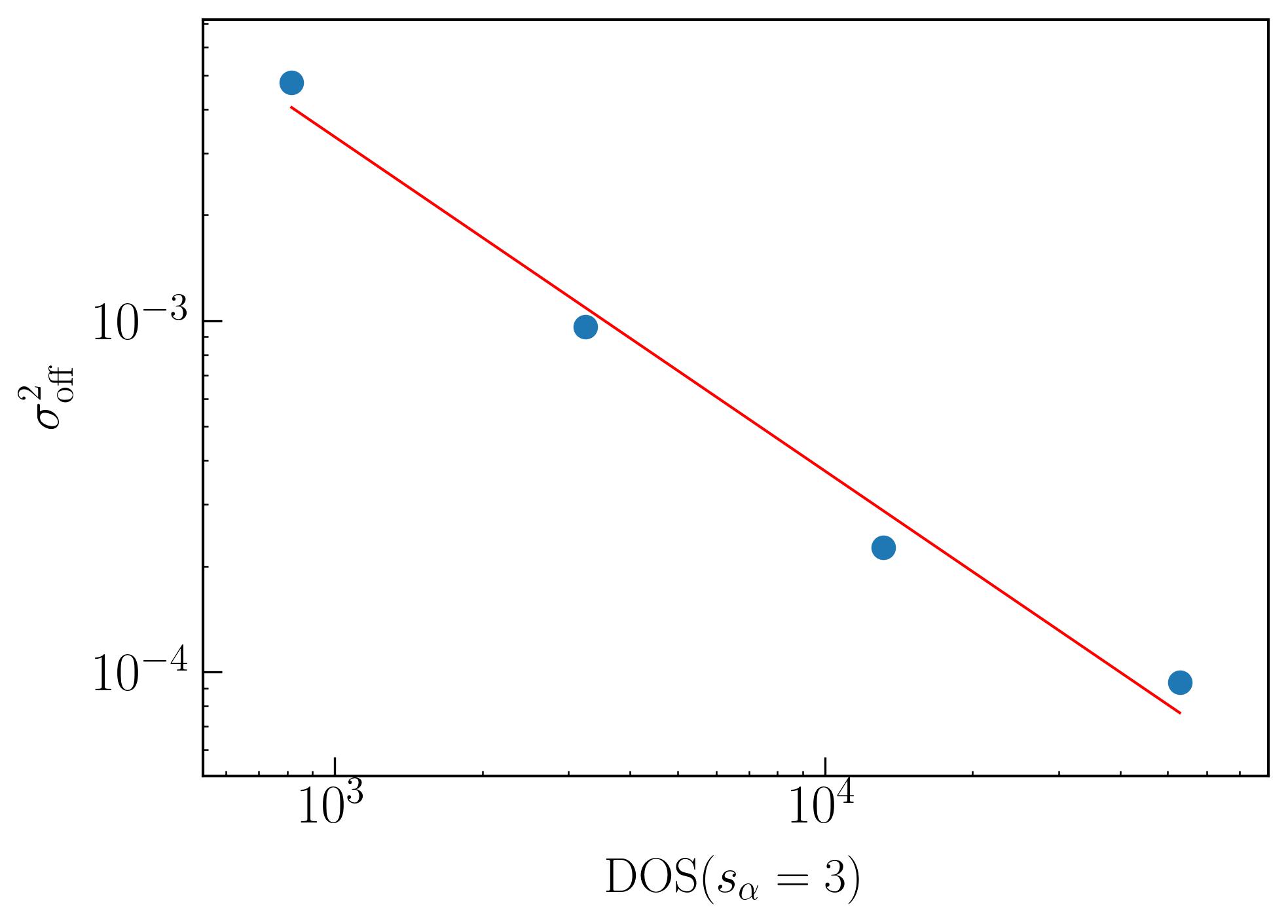}
    \caption{Off--block-diagonal reduced matrix elements' variance, $\sigma_{\rm off}^2$, versus DOS.  We sample the elements at $s_\alpha = s_{\alpha'} =3$ and such $\mathcal{E}$ that maximizes the DOS. 
    Data came from system sizes $N=12$–18. The log–log plot displays a linear relationship and negative slope. Hence the variance decays algebraically as the DOS grows. The slope is approximately $-0.95 \pm 0.09$; the non-Abelian ETH's final term predicts $-1$. The operator analyzed is $T^{(1)}_0$. 
    The linear relationship holds in the other parameter regimes checked. 
    }
    \label{fig:ExpDecay}
\end{figure}

Having shown empirically that the block-diagonal reduced matrix elements' variance decays exponentially as the entropy grows, we show that the off--block-diagonal reduced matrix elements' variance decays similarly. 
We sample $T^{(1)}_0$ elements labeled by $s_\alpha=s_{\alpha'}=3$. We draw from near (i) the $\mathcal{E}/N$ value that maximizes the DOS (Fig. \ref{fig:Q18_DOS}) and (ii)  $\omega=0$.
Figure~\ref{fig:ExpDecay} shows $\sigma_{\rm off}^2$ as a function of the DOS at system sizes $N=12$–18. The log–log plot's linear relationship implies that $\sigma^2_\mathrm{off}$ decays algebraically as the DOS grows. The slope, $ -0.95 \pm 0.09$, lies within a standard deviation of the predicted $-1$. 

Figure \ref{fig:ExpDecay} exhibits a slope closer to $-1$ than Fig. \ref{fig:ExpDecayDiag} does. The reason is that the off--block-diagonal effects are much less sparse than the block-diagonal elements and so suffer less from finite-size effects.

\section{Conclusions}
\label{sec_Conclusions}

We have extensively supported the non-Abelian ETH numerically. Our work centers on two Heisenberg models, which exhibit SU(2) symmetry. In the main text, we checked six qualitative predictions of the non-Abelian ETH (Figs.~\ref{fig:Q18_Energy_gaps_GEO}--\ref{fig:ffunctS_T10} and the $q$-independence reported in App.~\ref{app:data}) and one quantitative prediction (Fig.~\ref{fig:varratios_T10}). Additionally, we argued analytically that the non-Abelian ETH is self-consistent if the thermodynamic entropy $S_\th$ is smaller than anticipated in~\cite{MurthyNAETH}. These results offer the most comprehensive evidence, to date, for the non-Abelian ETH.

Our work establishes multiple opportunities for future research. First, the non-Abelian ETH implies how systems thermalize internally in the presence of non-Abelian symmetries~\cite{MurthyNAETH}. These predictions can now be checked; specific models' time evolutions should be simulated numerically and on small-scale quantum computers~\cite{NYH_20_Noncommuting,Kranzl_23_Experimental,NYH_22_How}. Our Heisenberg models offer starting points.

Second, our results encompass off--block-diagonal matrix elements, whereas~\cite{MurthyNAETH} focuses on block-diagonal elements. Leveraging the off--block-diagonal elements, we derive a fluctuation--dissipation relation~\cite{Noh_20_Numerical} for SU(2)-symmetric systems~\cite{25_Noh_Kubo}. One can also leverage the off--block-diagonal elements to ascertain how local operators' expectation values fluctuate about their time averages. Perhaps the non-Abelian symmetry causes deviations from conventional ETH-sourced results.

Second, the variables $R^{(T)}_{\alpha \alpha'}$ may be correlated. We demonstrated that $R^{(T)}_{\alpha \alpha'}$ values, viewed individually, are distributed in Gaussian fashion (Figs.~\ref{fig:Q18_Diag_Hist_s3} and~\ref{fig:Q18_Offdiag_Hist_s3}). We did not quantify correlations among the $R^{(T)}_{\alpha \alpha'}$s. Such correlations have recently undergone extensive study in conventional ETH contexts~\cite{Foini_19_Eigenstate,Jorge_2022_ETHprobability,Wang_2022_ETH_RMT}. The correlations affect long-time dynamics, Lyapunov exponents, and out-of-time-ordered correlators. Do non-Abelian symmetries change these results qualitatively? The mathematical toolkit of free probability theory could shed light on this question~\cite{Jorge_2022_ETHprobability}.

Third and most broadly, this work bridges noncommuting charges from quantum thermodynamics to many-body physics. This bridge has been expanding rapidly (e.g.,~\cite{NYH_20_Noncommuting,Zhang_20_Stationary,Fukai_20_Noncommutative,NYH_22_How,Kranzl_23_Experimental,MurthyNAETH,Majidy_23_Noncommuting,Marvian_23_Non,Corps_23_General,Majidy_23_Critical,Majidy_24_Noncommuting,Dabholkar_24_Ergodic,Li_23_Designs}), evidencing numerous opportunities. Examples include experimental tests of theoretical noncommuting-charge results. Another opportunity is to map out all the stages of many-body thermalization in the presence of non-Abelian symmetries, from local thermalization~\cite{MurthyNAETH} to information scrambling~\cite{Swingle_18_Unscrambling} to $k$-design generation~\cite{Li_23_Designs} to spectral form factors~\cite{Haake_10_Quantum} to quantum-complexity saturation~\cite{Brown_18_Second}.

\begin{acknowledgments}
J.~L. is grateful to the Institute for Quantum Computing, Ray Laflamme, and Shayan Majidy for their hospitality during the development of this paper.
N.~Y.~H. is grateful for conversations with Chaitanya Murthy and Mark Srednicki.
This work received support from the National Science Foundation 
(QLCI grant OMA-2120757 
and NSF award DMS-2231533),  
the John Templeton Foundation (award no. 62422),
a National Research Foundation of Korea~(NRF) grant funded by the Korea government (MSIT) (grant No. RS-2024-00348526), and 
a USEQUIP URA from Innovation, Science, and Economic Development (ISED) Canada. 
The opinions expressed in this publication are those of the authors and do not necessarily reflect the views of the John Templeton Foundation or UMD.
\end{acknowledgments}

\begin{appendices}

\onecolumngrid

\renewcommand{\thesection}{\Alph{section}}
\renewcommand{\thesubsection}{\Alph{section} \arabic{subsection}}
\renewcommand{\thesubsubsection}{\Alph{section} \arabic{subsection} \roman{subsubsection}}

\makeatletter\@addtoreset{equation}{section}
\def\theequation{\thesection\arabic{equation}}

\section{Self-consistency of the ETH}
\label{sec_Consistency}

Srednicki argued for the ETH's self-consistency by studying the square $A^2$ of an operator $A$~\cite{Srednicki_96_Thermal}. From the ETH, one can infer directly and indirectly how the matrix elements $(A^2)_{\alpha \alpha'}$ scale. The two scalings are the same, Srednicki confirmed. We recapitulate his argument here.

We reuse the setup in Sec.~\ref{sec_Backgrnd_ETH}. Additionally, we assume that the Hamiltonian $H$ and local operator $A$ obey the ETH [Eq.~\eqref{eq_ETH}]. Denote by $\Dim_\mathcal{E}$ the DOS at $\mathcal{E}$.
The thermodynamic entropy $S_{\rm th} ( \mathcal{E} ) = \log(\Dim_\mathcal{E})$. Denote by $D$ the global Hilbert space's dimensionality.

$A^2$ is an operator similarly to $A$. Like the matrix elements $A_{\alpha \alpha'}$, the elements $( A^2 )_{\alpha \alpha'}$ can directly obey the ETH---can satisfy Eq.~\eqref{eq_ETH}. Also, because the $A_{\alpha \alpha'}$s satisfy Eq.~\eqref{eq_ETH}, the ETH indirectly determines the structure of
\begin{align}
   \label{eq_ETH_Decompose}
   \left( A^2  \right)_{\alpha \alpha'}  
   =  \sum_{\beta = 1}^D  
   A_{\alpha \beta}  \,  A_{\beta \alpha'} \, ,
\end{align}
by determining each factor's structure. 
We expound upon the direct implication, then the indirect implication. Afterward, we compare the two.

\textbf{Direct implication of the ETH:} Let us substitute $A^2$ into the ETH [Eq.~\eqref{eq_ETH}]:
\begin{align}
   \label{eq_ETH2}
   \left( A^2 \right)_{\alpha \alpha'}
   & =  \mathcal{A}_2 \left( \mathcal{E} \right)  \,  \delta_{\alpha \alpha'}
   +  e^{ - S_{\rm th} \left( \mathcal{E} \right) / 2 }  \,
   f^{ ( A^2 ) }  \left( \mathcal{E}, \omega \right)  \,
   R^{( A^2)}_{\alpha \alpha'} .
\end{align}
$\mathcal{A}_2$ and $f^{ ( A^2 ) }$ denote smooth, real, $O(1)$ functions.
$R^{( A^2)}_{\alpha \alpha'}$ varies as $R_{\alpha \alpha'}$ does.
On the block diagonal (when $\alpha' = \alpha$), therefore, the matrix element scales as
\begin{align}
   \label{eq_Direct_Scale_Diag}
   \left( A^2 \right)_{\alpha \alpha}
   =  O (1)  +  O \left(  \Dim_\mathcal{E}^{-1/2}  \right)  .
\end{align}
Off the block diagonal, the matrix element scales as
\begin{align}
   \label{eq_Direct_Scale_Offdiag}
   \left( A^2 \right)_{\alpha \alpha'}
   =  0  +  O \left(  \Dim_\mathcal{E}^{-1/2}  \right)
   =  O \left(  \Dim_\mathcal{E}^{-1/2}  \right) .
\end{align}
Recalling the terms' sources is worthwhile. In Eqs.~\eqref{eq_Direct_Scale_Diag} and~\eqref{eq_Direct_Scale_Offdiag}, the 
$O \left(  \Dim_\mathcal{E}^{-1/2}  \right)$ terms come from random numbers whose signs fluctuate. The $O(1)$ term comes from a deterministic function.

\textbf{Indirect implication of the ETH:} Equation~\eqref{eq_ETH_Decompose} shows how $( A^2 )_{\alpha \alpha'}$ decomposes in terms of $A$ matrix elements. Let us substitute in for those elements from the ETH [Eq.~\eqref{eq_ETH}].
We must introduce new definitions:
$\mathcal{E}_{\gamma \epsilon}  \coloneqq  ( E_\gamma + E_\epsilon ) / 2$, and
$\omega_{\gamma \epsilon}  \coloneqq  E_\gamma - E_\epsilon \, .$
We multiply out in Eq.~\eqref{eq_ETH_Decompose} and evaluate the Kronecker deltas:
\begin{align}
   \label{eq_ETH_Indirect1}
   \left( A^2  \right)_{\alpha \alpha'}
   & =  \mathcal{A}^2  \left( E_\alpha  \right)  \,  \delta_{\alpha \alpha'}
   +  \left[ \mathcal{A}  \left( E_\alpha \right)  
               +  \mathcal{A} \left( E_{\alpha'} \right) \right]  \,
       e^{ - S_{\rm th} \left( \mathcal{E} \right) / 2 }  \,
       f \left( \mathcal{E}, \omega \right)  \,  R_{\alpha \alpha'}
   \\ & \nonumber \quad
   +  \sum_{\beta = 1}^D  e^{ - \left[ S_{\rm th} ( \mathcal{E}_{\alpha \beta} )  
   +  S_{\rm th} ( \mathcal{E}_{\beta \alpha'} )  \right] / 2 }  \,
   f \left( \mathcal{E}_{\alpha \beta},  \omega_{\alpha \beta} \right)  \,
   f \big( \mathcal{E}_{\beta \alpha'} \, , \omega_{\beta \alpha'} \big)  \,
   R_{\alpha \beta} \,  R_{\beta \alpha'} .
\end{align}

We now evaluate how each term scales.
The first term is $O(1)$ on the block diagonal and vanishes off the block diagonal.
The second term is $O \big( \Dim_\mathcal{E}^{-1/2} \big)$.

The third term requires more–in-depth analysis. Srednicki does not address the thermodynamic entropy's or the $f$ functions' $\beta$-dependence. (Relatedly, he does not separate the $f$ functions from the $R$s.) We therefore augment his argument, following page 315 of~\cite{DAlessio_16_From}. Each $f$ function decays rapidly as $| \omega |$ grows (Sec.~\ref{sec_results}). In the third term of Eq.~\eqref{eq_ETH_Indirect1}, therefore, the summand is non-negligible only if $E_\beta \approx E_\alpha \, , E_{\alpha'} \, .$ The number of such $\beta$ values is $O( \Dim_\mathcal{E} )$. Furthermore, $S_\th$ changes significantly only if its argument changes extensively. We therefore approximate both entropies at $\mathcal{E}$:
\begin{align}
   \label{eq_ETH_Indirect2}
   \sum_{\beta = 1}^D 
   e^{ - \left[ S_{\rm th} ( \mathcal{E}_{\alpha \beta} )  
   +  S_{\rm th} ( \mathcal{E}_{\beta \alpha'} )  \right] / 2 }  \,
   f \left( \mathcal{E}_{\alpha \beta} \, , \omega_{\alpha \beta} \right)  \,
   f \big( \mathcal{E}_{\beta \alpha'} \, , \omega_{\beta \alpha'} \big)  \,
   R_{\alpha \beta} \,  R_{\beta \alpha'}
   \approx e^{ -S_\th (\mathcal{E}) }
   \sum_{\beta=1}^{ O( \Dim_\mathcal{E} ) } 
   f \left( \mathcal{E}_{\alpha \beta} \, , \omega_{\alpha \beta} \right)  \,
   f \big( \mathcal{E}_{\beta \alpha'} \, , \omega_{\beta \alpha'} \big)  \,
   R_{\alpha \beta} \,  R_{\beta \alpha'} \, .
\end{align}
The exponential factor equals $\Dim_\mathcal{E}^{-1}$.

Suppose that $\alpha' = \alpha$. Imagine temporarily that the $f$ functions were absent. $| R_{\alpha \beta} |^2$ fluctuates randomly about its mean as a function of $\beta$. The fluctuations would approximately cancel out in
$\sum_{\beta=1}^{ O( \Dim_\mathcal{E} ) } 
| R_{\alpha \beta} |^2$.
We could therefore approximate the $| R_{\alpha \beta} |^2$ with its average value, which is $O(1)$. Hence
$\sum_{\beta=1}^{ O( \Dim_\mathcal{E} ) } 
| R_{\alpha \beta} |^2
= O(\Dim_\mathcal{E})$.
Now, consider the $f$ functions. They are smooth, $O(1)$ functions when $E_\beta \approx E_{\alpha} \, , E_{\alpha'} \, .$ Therefore, they only cause the summand's mean to trend slowly with $\beta$; they do not alter our conclusion. That is, if $\alpha' = \alpha$, then the third term in Eq.~\eqref{eq_ETH_Indirect1} is
\begin{align}
   O \left( \Dim_\mathcal{E}^{-1} \right) \times 
   O ( \Dim_\mathcal{E} )
   = O (1).
\end{align}

Now, suppose that $\alpha' \neq \alpha$. Each $R$ factor has a random sign; so $R_{\alpha \beta} \,  R_{\beta \alpha'}$ does, as well. Hence 
$\sum_{\beta=1}^{ O( \Dim_\mathcal{E} ) } 
R_{\alpha \beta} \,  R_{\beta \alpha'}$
approximates the typical displacement of a random walker
after $\Dim_\mathcal{E}$ steps, 
$O \left(  \Dim_\mathcal{E}^{1/2}  \right)$.
This conclusion remains invariant if we introduce the smooth $f$ functions into the summand. Hence the third term in Eq.~\eqref{eq_ETH_Indirect2} is 
\begin{align}
   O \left( \Dim_\mathcal{E}^{-1} \right) \times
   O \left(  \Dim_\mathcal{E}^{1/2}  \right)
   = O \left(  \Dim_\mathcal{E}^{-1/2}  \right) 
\end{align}
if $\alpha' \neq \alpha$.

Let us combine all this scaling analysis with Eq.~\eqref{eq_ETH_Indirect1}. On the block diagonal,
\begin{align}
   \label{eq_A2_Indirect_Diag}
   \left( A^2 \right)_{\alpha \alpha}
   =  O (1)  +  O \left( \Dim_\mathcal{E}^{-1/2}  \right)  +  O(1)
   =  O(1)  +  O \left( \Dim_\mathcal{E}^{-1/2}  \right) .
\end{align}
In the central expression, the final two terms come from randomly fluctuating numbers. Off the block diagonal,
\begin{align}
   \label{eq_A2_Indirect_Offdiag}
   \left( A^2 \right)_{\alpha \alpha'}
   =  0  +  O \left( \Dim_\mathcal{E}^{-1/2}  \right)  
      +  O \left( \Dim_\mathcal{E}^{-1/2}  \right)
   =  O \left( \Dim_\mathcal{E}^{-1/2}  \right) .
\end{align}
All contributions come from randomly fluctuating numbers.

\textbf{Comparison:} The direct and indirect derivations agree about how the block-diagonal matrix elements $( A^2 )_{\alpha \alpha}$ scale, according to Eqs.~\eqref{eq_Direct_Scale_Diag} and~\eqref{eq_A2_Indirect_Diag}.
According to Eqs.~\eqref{eq_Direct_Scale_Offdiag} and~\eqref{eq_A2_Indirect_Offdiag}, the derivations agree about the off--block-diagonal matrix elements $( A^2 )_{\alpha \alpha'}$.

\section{Detailed argument for the non-Abelian ETH's self-consistency}
\label{app_2s}

Section~\ref{sec_2s_main} sketches this argument and details part of it. The argument concerns a spherical tensor operator $T^\KParen_q$, which decomposes in accordance with Eq.~\eqref{eq:combineAB}:
\begin{align}
   \label{eq_combine_AB_app}
   T_{q}^{(k)}
   =\sum_{q^{\prime}}\left\langle k, q \mid k_{1}, q^{\prime} ; k_{2}, q-q^{\prime}\right\rangle 
   A_{q^{\prime}}^{\left(k_{1}\right)} B_{q-q^{\prime}}^{\left(k_{2}\right)}\,.
\end{align}
The non-Abelian ETH directly implies how the block-diagonal reduced matrix element 
$\langle \alpha || T^\KParen || \alpha \rangle$ scales, as shown in Eq.~\eqref{eq_NAETH_Direct_Scale}:
\begin{align}
   \label{eq_NAETH_Direct_Scale_app}
   \langle \alpha || T^\KParen || \alpha \rangle
   =  O (1)  +  O \big( \Dim^{-1/2}  \big) .
\end{align}
Recall that we defined $\Dim \coloneqq e^{S_\th(\mathcal{E}, \mathcal{S}) }$. In App.~\ref{app_Self_cons_Diag}, we infer the non-Abelian ETH's \emph{indirect} implication for how $\langle \alpha || T^\KParen || \alpha \rangle$ scales. This implication must have the same form as Eq.~\eqref{eq_NAETH_Direct_Scale_app}, for the non-Abelian ETH to be self-consistent. The non-Abelian ETH meets this condition if the thermodynamic entropy $S_\th (\mathcal{E}, \mathcal{S})$ is defined as at the end of Sec.~\ref{sec_Backgrnd_NAETH}. Appendix~\ref{app_Self_cons_Off_diag} extends this self-consistency argument to off--block-diagonal matrix elements
$\langle \alpha || T^\KParen || \alpha' \rangle$, wherein $\alpha' \neq \alpha$.

\subsection{Self-consistency of the non-Abelian ETH's predictions about block-diagonal reduced matrix elements}
\label{app_Self_cons_Diag}

Let us infer from the $T^\KParen_q$ decomposition~\eqref{eq_combine_AB_app} about how 
$\langle \alpha || T^\KParen || \alpha' \rangle$ scales. (The beginning of this argument is not specific to block-diagonal matrix elements.) We operate on the equation's LHS with $\bra{\alpha, m}$ and on the RHS with $\ket{\alpha', m'}$. We also insert a resolution of unity,
$\id = \sum_{\alpha'', \, m''}  \ketbra{\alpha'', m''}{\alpha'', m''}$,
between the $A$ and $B$ operators. The $\sum_{\alpha''}$ runs over the number of $s_{\alpha''}$ values, which is $O(N)$. Next, we apply the Wigner–Eckart theorem to each side of the equation:
\begin{align}
   \label{eq_2s_help1}
   \braket{s_\alpha \, , m}{ s_{\alpha'} \, , m' \, ; k, q}
   \langle \alpha || T^\KParen  ||  \alpha' \rangle 
   & =  \sum_{q'}  \braket{k, q}{ k_1 \, , q' \, ; k_2 \, , q - q' }
   \sum_{\alpha'' \, , m''}
   \braket{s_\alpha , m}{ s_{\alpha''} \, , m'' \, ; k_1 \, , q'}
   \\ & \qquad \quad \nonumber \times
   \langle \alpha || A^{(k_1)} || \alpha'' \rangle
   \braket{ s_{\alpha''} \, , m'' }{ s_{\alpha'} \, , m' \, ; k_2 \, , q - q' }
   \langle \alpha'' || B^{(k_2)} || \alpha' \rangle .
\end{align}
We now invoke a Clebsch–Gordan property that underlies selection rules in AMO physics: 
$\braket{ s_\alpha , m }{ s_{\alpha''} \, , m'' \, ; k_1 \, , q' }$ 
is nonzero only if $m'' = m - q'$.
Similarly, $\braket{ s_{\alpha''} \, , m'' }{ s_{\alpha'} \, , m' \, ; k_2 \, , q - q' }$
is nonzero only if $m' = m'' - (q - q')$.
Combining these two implications yields $m'' = m - q'$ and $m' = m - q$.
Using these relations, we eliminate $m''$ and $m'$ from Eq.~\eqref{eq_2s_help1}:
\begin{align}
   \label{eq_2s_help2} 
   \braket{s_\alpha \, , m}{ s_{\alpha'} \, , m - q \, ; k, q}
   \langle \alpha || T^\KParen  ||  \alpha' \rangle
&   =  \sum_{q'} \langle k, q | k_1 \, , q' \, ; k_2 \, , q - q' \rangle
   \sum_{\alpha''}  \braket{ s_\alpha , m }{s_{\alpha''} \, , m - q' \, ; k_1 \, , q' }
   \langle \alpha || A^{(k_1)} || \alpha'' \rangle
   \nonumber \\ & \qquad \quad \times
   \braket{ s_{\alpha''} \, , m - q' }{ s_{\alpha'} \, , m - q; k_2 \, , q - q' }
   \langle \alpha''  ||  B^{(k_2)}  ||  \alpha' \rangle .
\end{align}

We solve for $\langle \alpha || T^\KParen  ||  \alpha' \rangle$, assuming that its prefactor is nonzero. The equation's RHS comes to contain all the Clebsch–Gordan coefficients, consolidated in
\begin{align}
   \label{eq_Ups_1}
   & \Upsilon ( s_\alpha , s_{\alpha'} \, , s_{\alpha''} \, , k, k_1 \, , k_2 \, , m, q ) 
   \\ \nonumber
   & \coloneqq  \sum_{q'}  
   \langle k, q | k_1 \, , q' \, ; k_2 \, , q - q' \rangle
   \braket{ s_\alpha , m }{s_{\alpha''} \, , m - q' \, ; k_1 \, , q' }\braket{ s_{\alpha''} \, , m - q' }{ s_{\alpha'} \, , m - q; k_2 \, , q - q' }
   /  \braket{s_\alpha \, , m}{ s_{\alpha'} \, , m - q \, ; k, q} .
\end{align}
For notational simplicity, we now suppress $k$-type indexes and arguments. Combining Eqs.~\eqref{eq_Ups_1} and~\eqref{eq_2s_help2} yields
\begin{align}
   \label{eq_2s_help3}
   \langle \alpha || T^\KParen ||  \alpha' \rangle
   =  \sum_{\alpha''}
   \Upsilon ( s_\alpha , s_{\alpha'} \, , s_{\alpha''} \, , m, q ) \,
   \langle \alpha || A || \alpha'' \rangle
   \langle \alpha''  ||  B  ||  \alpha' \rangle .
\end{align}
$\Upsilon$ is the only factor that can depend on $m$ and $q$. Therefore, $\Upsilon$ must not depend on these arguments. The Wigner–Eckart theorem underlies this conclusion. We therefore redefine the symbol:
$\Upsilon ( s_\alpha , s_{\alpha'} \, , s_{\alpha''} \, , m, q )
\equiv  \Upsilon ( s_\alpha , s_{\alpha'} \, , s_{\alpha''} )$.

Let us assume that $A$ and $B$ obey the non-Abelian ETH. We define 
$\mathcal{E}_{\beta \gamma} \coloneqq (E_\beta + E_\gamma) / 2$, 
$\mathcal{S}_{\beta \gamma} \coloneqq (s_\beta + s_\gamma)/2$,
$\omega_{\beta \gamma} \coloneqq E_\beta - E_\gamma$, and $\nu_{\beta \gamma} \coloneqq s_\beta - s_\gamma$.
Our assumption has the mathematical form
\begin{align}
   \label{eq_A_NAETH}
   & \langle \alpha || A || \alpha'' \rangle
   =  \mathcal{A} \left( \mathcal{E}_{\alpha \alpha''} \, ,  \mathcal{S}_{\alpha \alpha''}  \right) \,
   \delta_{\alpha \alpha''}
   +  e^{- S_\th  ( \mathcal{E}_{\alpha \alpha''} \, , \, \mathcal{S}_{\alpha \alpha''}  ) / 2} \,
   f^{(A)}_{ \nu_{\alpha \alpha''} }  
   \left( \mathcal{E}_{\alpha \alpha''} \, , \,  \mathcal{S}_{\alpha \alpha''} \, , \omega_{\alpha \alpha''}  \right)  \,
   R^{(A)}_{\alpha \alpha''}  
   \quad \text{and} \\
   \label{eq_B_NAETH}
  & \langle \alpha''  ||  B  ||  \alpha' \rangle 
   =  \mathcal{B} \left( \mathcal{E}_{\alpha'' \alpha'} \, , \mathcal{S}_{\alpha'' \alpha'}  \right) \,
   \delta_{\alpha'' \alpha'}
   +  e^{- S_\th ( \mathcal{E}_{\alpha'' \alpha'} \, , \, \mathcal{S}_{\alpha'' \alpha'} ) / 2}  \,
   f^{(B)}_{ \nu_{\alpha'' \alpha'} }  
   \left(  \mathcal{E}_{\alpha'' \alpha'} \, , \, \mathcal{S}_{\alpha'' \alpha'} \, ,  \omega_{\alpha'' \alpha'}  \right)  \,
   R^{(B)}_{\alpha'' \alpha'} \, .
\end{align}
The $\mathcal{A}$, $\mathcal{B}$, $f$, and $R$ quantities behave like their analogues in Sec.~\ref{sec_Backgrnd_NAETH}.
We substitute from Eqs.~\eqref{eq_A_NAETH} and~\eqref{eq_B_NAETH} into the reduced-matrix-element equation~\eqref{eq_2s_help3}, focusing on a block-diagonal element. We multiply out terms and evaluate the Kronecker delta functions:
\begin{align}
   \label{eq_2s_help4}
   \langle \alpha || T  ||  \alpha \rangle
   & =  \Upsilon ( s_\alpha, s_\alpha, s_\alpha )
   \big\{  \mathcal{A} \left ( \mathcal{E}_{\alpha \alpha} ,  \mathcal{S}_{\alpha \alpha}  \right) \,
             \mathcal{B}  \left ( \mathcal{E}_{\alpha \alpha} ,  \mathcal{S}_{\alpha \alpha}  \right)
   \\ \nonumber & \quad
   +  e^{- S_\th ( \mathcal{E}_{\alpha \alpha} ,  \mathcal{S}_{\alpha \alpha} ) / 2 }
       \left[  \mathcal{A} \left ( \mathcal{E}_{\alpha \alpha} ,  \mathcal{S}_{\alpha \alpha}  \right)
             f^{(B)}_{ \nu_{\alpha \alpha} }  \left(  \mathcal{E}_{\alpha \alpha} ,  \mathcal{S}_{\alpha \alpha} ,  0  \right)  \,
             R^{(B)}_{\alpha \alpha}
             +  \mathcal{B}  \left( \mathcal{E}_{\alpha \alpha} , \mathcal{S}_{\alpha \alpha}  \right) \,
             f^{(A)}_{ \nu_{\alpha \alpha} }  \left(  \mathcal{E}_{\alpha \alpha} ,  \mathcal{S}_{\alpha \alpha} ,  0  \right)  \,
             R^{(A)}_{\alpha \alpha}
       \right]  \Big\} 
       \\ \nonumber & \quad
   +  \sum_{\alpha''}  \Upsilon (s_\alpha ,  s_\alpha,  s_{\alpha''} )  \,
   e^{- S_\th ( \mathcal{E}_{\alpha \alpha''} \, , \,  \mathcal{S}_{\alpha \alpha''} ) }  \,
   f^{(A)}_{ \nu_{\alpha \alpha''} }  \left( \mathcal{E}_{\alpha \alpha''} \, ,  \mathcal{S}_{\alpha \alpha''} \, ,  \omega_{\alpha \alpha''}  \right)  \,
   f^{(B)}_{ \nu_{\alpha'' \alpha}  }  
   \left( \mathcal{E}_{\alpha'' \alpha} \, ,  \mathcal{S}_{\alpha'' \alpha} \, ,  \omega_{\alpha'' \alpha}  \right)  \,
   R^{(A)}_{\alpha \alpha''}  \,
   R^{(B)}_{\alpha'' \alpha}  \, .
\end{align}

We now simplify and approximate $\Upsilon$. Below Eq.~\eqref{eq_2s_help3}, we concluded that $\Upsilon$ cannot depend on $m$ or $q$. Therefore, we can evaluate $\Upsilon$ [using Eq.~\eqref{eq_Ups_1}] at arbitrary $m$ and $q$ values to infer about the function's general behavior.\footnote{
This $m$ value labels the energy eigenstate $\ket{\alpha, m}$ used to define the matrix element 
$\bra{\alpha, m} T^{(k)}_q \ket{\alpha, m}$ (top of App.~\ref{app_Self_cons_Diag}). This $m$ value plays no role in the DOS that appears implicitly in the non-Abelian ETH (below Eq.~\eqref{eq_NAETH}): 
$S_\th (\mathcal{E}, \mathcal{S}) = \log(d)$. No $m$ appears in this equation. Rather, we defined $d$ as the DOS that the system would have at $(\mathcal{E}, \mathcal{S})$ if it lacked any degeneracy attributable to the magnetic spin quantum number. This $d$ definition ensures the non-Abelian ETH's self-consistency.}
We choose $m = s_\alpha$ and $q = s_\alpha - s_{\alpha'}$, for ease of evaluation. Equation~\eqref{eq_Ups_1} simplifies to
\begin{align}
   \label{eq_Ups_2}
   \Upsilon ( s_\alpha , s_{\alpha'} \, , s_{\alpha''} )
   & =  \sum_{q'}  \braket{k,  s_\alpha - s_{\alpha'} }{ k_1 \, , q' ;  k_2 ,  s_\alpha - s_{\alpha'} - q'}
   \braket{ s_\alpha , s_\alpha }{ s_{\alpha''} \, , s_\alpha - q' \, ; k_1 \, , q' }
   \\ \nonumber & \qquad \quad \times 
   \braket{ s_{\alpha''} \, , s_\alpha - q' }{ s_{\alpha'} \, , s_{\alpha'} \, ; k_2 \, , s_\alpha - s_{\alpha'} - q' }
   /  \braket{ s_\alpha , s_\alpha }{ s_{\alpha'} \, , s_{\alpha'} \, ; k, s_\alpha - s_{\alpha'} } .
\end{align}

The foregoing formula contains four Clebsch–Gordan coefficients, three of which we now approximate.\footnote{
We need not approximate if 
$s_\alpha, s_{\alpha'}, s_{\alpha''} = O(1)$. Under this condition, $\Upsilon$ depends only on $O(1)$ numbers and so is $O(1)$. The result $\Upsilon = O(1)$ underlies this appendix's main result.}
In App.~\ref{sec_CG_approx}, we prove that
\begin{align}
   \label{eq_CG_approx} &
   \braket{s + \nu , m + q }{ s, m; k, q}
   = c (\nu, k, q, s - m ) \,
   s^{- | \nu - q | / 2}  \,
   \left[ 1 + O \left(s^{-1} \right) \right] 
   \quad \text{if} \\
   & \label{eq_CG_approx_assume0}
   s, m = O(N^\delta) , 
   \; \; \text{wherein} \; \; \delta \in (0, \, 1] , 
   \; \; \text{and} \;  \;
   \nu, k, q, s-m = O(N^0) .
\end{align}
The function $c$ equals 1 if $\nu = q$ and is $O(1)$ regardless. We can understand the condition $s, m \gg \nu, k , q, s - m$ physically in terms of the atom-and-photon example from Sec.~\ref{sec_Backgrnd_Oprs}. The atom must begin with a great deal of angular momentum, the photon must begin with little, and the photon must impart little to the atom.

We now apply the Clebsch–Gordan approximation~\eqref{eq_CG_approx} to $\Upsilon$ [Eq.~\eqref{eq_Ups_2}]. To do so, we assume that all the spin quantum numbers have similar, large values:
\begin{align}
   \label{eq_s_Large} &
   s_\alpha , s_{\alpha'} \, , s_{\alpha''}
   = O \left( N^\delta \right) ,
   \;  \; \text{wherein} \;  \; 
   \delta \in (0, 1] ,
   \;  \; \text{and} \\ &
   \label{eq_s_Large_2}
   | s_\alpha - s_{\alpha''} | \, , \, 
   | s_{\alpha'} - s_{\alpha''} |
   = O(1) .
\end{align}
Equation~\eqref{eq_s_Large} is consistent with ETHs' tendency to govern matrix elements defined in terms of Hamiltonian eigenstates associated with large energies, which tend to accompany large $s_\alpha$ values. Also, Eq.~\eqref{eq_s_Large} is necessary for distinguishing, via a scaling argument, whether the non-Abelian ETH's $S_\th$ should be defined as in~\cite{MurthyNAETH} or as at the end of Sec.~\ref{sec_Backgrnd_NAETH}. The two options differ by a $2 \mathcal{S} + 1$ factor, as explained later in this appendix. If all the relevant $s$ values were $O(1)$, then the factor would be $O(1)$. A scaling analysis therefore could not distinguish whether the factor should be present or absent. 
Having justified the assumption~\eqref{eq_s_Large}, we return to $\Upsilon$ [Eq.~\eqref{eq_Ups_2}]. Of the Clebsch–Gordan coefficients in $\Upsilon$, which obey the conditions~\eqref{eq_CG_approx}--\eqref{eq_s_Large_2}? The first coefficient does not, but the other coefficients do:
\begin{align}
   \label{eq_CG_approx_ex_1}
   & \braket{ s_\alpha , s_\alpha }{ s_{\alpha''} \, , s_\alpha - q' \, ; k_1 \, , q'}
   =  c  \left( s_\alpha - s_{\alpha''} \, , k_1 \, , q' , s_{\alpha''} 
                   -  \left[ s_\alpha - q'  \right]  \right) \,
   s_{\alpha''}^{- | s_\alpha - s_{\alpha''} - q' | / 2}  \,
   \left[ 1  +  O \left( s_{\alpha''}^{-1}  \right)  \right] , \\
   \label{eq_CG_approx_ex_2}
   & \braket{ s_{\alpha''} \, , s_\alpha - q' }{ s_{\alpha'} \, , s_{\alpha'} \, ; k_2 , s_\alpha - s_{\alpha'} - q' }
   =  c ( s_{\alpha''} - s_{\alpha'} \, , k_2 , s_\alpha - s_{\alpha'} - q' , 0 ) \,
   s_{\alpha'}^{- | q' - (s_\alpha - s_{\alpha''} ) | / 2} \, 
   \left[ 1 + O \left( s_{\alpha'}^{-1} \right) \right] \, , 
   \; \text{and} \\
   \label{eq_CG_approx_ex_3}
   & \braket{ s_\alpha , s_\alpha }{ s_{\alpha'} \, , s_{\alpha'} \, ; k, s_\alpha - s_{\alpha'} } \,
   =  
   1 + O \left( s_{\alpha'}^{-1}  \right) .
\end{align}
These formulae imply which terms in the $\Upsilon$ formula~\eqref{eq_Ups_2} contribute significantly to the sum: the terms in which $q' = s_\alpha - s_{\alpha''}$. All other terms are $O( N^{-\delta} )$. We therefore assume, from now on, that $q' = s_\alpha - s_{\alpha''}$. In Eqs.~\eqref{eq_CG_approx_ex_1} and~\eqref{eq_CG_approx_ex_2}, the $c$ functions evaluate to 1. Substituting into the $\Upsilon$ formula~\eqref{eq_Ups_2} yields
\begin{align}
   \label{eq_Ups_3}
   \Upsilon ( s_\alpha , s_{\alpha'} \, , s_{\alpha''} )
   & =  \braket{ k, s_\alpha - s_{\alpha'} }{ k_1 \, , s_\alpha - s_{\alpha''} \, ; k_2 , s_{\alpha''} - s_{\alpha'} }
   \left[ 1 + O \left( s_{\alpha''}^{-1} \right) \right]
   \left[ 1 + O \left( s_{\alpha'}^{-1} \right) \right]^2 \\
   & = \braket{ k, s_\alpha - s_{\alpha'} }{ k_1 \, , s_\alpha - s_{\alpha''} \, ; k_2 , s_{\alpha''} - s_{\alpha'} }
   +  O  \left( N^{-\delta}  \right) .
\end{align}
The Clebsch--Gordan coefficient, depending only on $O(1)$ numbers, is $O(1)$. The $O (N^{-\delta})$ correction is negligible in comparison. We therefore drop the correction later, although we include it for completeness for now. We substitute for $\Upsilon$ into the non-Abelian ETH's indirect implication, Eq.~\eqref{eq_2s_help4}:
\begin{align}
   \label{eq_2s_help5} 
   \langle \alpha || T || \alpha \rangle
   & = \left[ \braket{ k, 0 }{ k_1 \, , 0; k_2 , 0}  
   +  O  \left( N^{-\delta}  \right) \right]
   \big\{ \mathcal{A} \left( \mathcal{E}_{\alpha\alpha} , \mathcal{S}_{\alpha\alpha} \right) \,
   \mathcal{B}  \left(  \mathcal{E}_{\alpha\alpha} ,  \mathcal{S}_{\alpha\alpha}  \right) 
   \\ \nonumber & \quad \;
   +  e^{- S_\th ( \mathcal{E}_{\alpha\alpha} , \mathcal{S}_{\alpha\alpha} ) / 2 } \,
   \Big[ \mathcal{A}  \left( \mathcal{E}_{\alpha\alpha} , \mathcal{S}_{\alpha\alpha}  \right) \,
   f^{(B)}_{\nu_{\alpha\alpha}}  
   \left( \mathcal{E}_{\alpha\alpha} , \mathcal{S}_{\alpha\alpha} , 0 \right) \, 
   R^{(B)}_{\alpha\alpha}
   +  \mathcal{B}  \left( \mathcal{E}_{\alpha\alpha} , \mathcal{S}_{\alpha\alpha}  \right) \,
   f^{(A)}_{\nu_{\alpha\alpha}}  \left(  \mathcal{E}_{\alpha\alpha} , \mathcal{S}_{\alpha\alpha} , 0 \right)  \,
   R^{(A)}_{\alpha\alpha}  \Big]  \Big\} 
   \\ \nonumber & \quad
   +  \sum_{\alpha''}
   \left[ \braket{k, 0}{ k_1 \, , s_\alpha - s_{\alpha''} \, ; k_2 , s_{\alpha''} - s_\alpha } 
          + O  \left( N^{-\delta}  \right) \right]
   e^{ - S_\th ( \mathcal{E}_{\alpha \alpha''} \, , \, \mathcal{S}_{\alpha \alpha''} ) } \,
   \\ \nonumber & \qquad \qquad \times
   f^{(A)}_{ \nu_{\alpha \alpha''} }  \left(  \mathcal{E}_{\alpha \alpha''} \, , \mathcal{S}_{\alpha \alpha''} \, , \omega_{\alpha \alpha''}  \right)  \,
   f^{(B)}_{ \nu_{\alpha'' \alpha} }  
   \left(  \mathcal{E}_{\alpha'' \alpha} \, , \mathcal{S}_{\alpha'' \alpha} \, , \omega_{\alpha'' \alpha}  \right) \,
   R^{(A)}_{\alpha \alpha''}  \,
   R^{(B)}_{\alpha'' \alpha } \, .
\end{align}

Let us assess each term's scaling with $N$. The Clebsch–Gordan coefficient $ \braket{ k, 0 }{ k_1 \, , 0; k_2 , 0}$ depends only on $O(1)$ arguments and hence is $O(1)$. So are $\mathcal{A}$, $\mathcal{B}$, the $f$s, and the $R$s, by Sec.~\ref{sec_Backgrnd_NAETH}. Therefore, the first term in Eq.~\eqref{eq_2s_help5}---the term that contains curly braces---is
\begin{align}
   \label{eq_NAETH_scale_help0}
   \left[ O(1) + O \left( N^{-\delta} \right) \right]
   \left[ O(1) + O \left( \Dim^{-1/2} \right) \times O(1) \right]
   = O(1) + O \left( \Dim^{-1/2} \right) .
\end{align}
As explained below Eq.~\eqref{eq_Ups_3}, we have dropped the $O(N^{-\delta})$ correction: it is negligible compared to the $O(1)$ term that it corrects. The $O(1)$ term comes from deterministic functions, whereas random variables contribute to the $O ( \Dim^{-1/2} )$.

Now, we assess how the final term in Eq.~\eqref{eq_2s_help5} scales. Each $f$ function decays rapidly as $| \omega_{\alpha \alpha''} |$ grows (Sec.~\ref{sec_results}). Therefore, we can approximate the $\sum_{\alpha''}$ by truncating it. The true $\sum_{\alpha''}$ runs over all the $s_{\alpha''}$ values, as noted above Eq.~\eqref{eq_2s_help1}. The summands are non-negligible only when $E_{\alpha''} \approx E_\alpha$. The number of such $E_{\alpha''}$ values would roughly equal the density of states at $E_\alpha = \mathcal{E}$ and $S_\alpha = \mathcal{S}$, if the sum were over both $\alpha''$ and $m''$. However, a Clebsch–Gordan coefficient eliminated the $\sum_{m''}$ in Eq.~\eqref{eq_2s_help1}. If the $\sum_{m''}$ were still present, it would contribute a $2 s_{\alpha''} + 1$ factor to the number of relevant $E_{\alpha''}$ values. In the absence of the $\sum_{m''}$, therefore, the number of relevant $E_{\alpha''}$ values---the number of non-negligible terms in the $\sum_{\alpha''}$---approximately equals the DOS at $\mathcal{E}$, $\mathcal{S}$, and magnetic spin quantum number 0 (if $N$ is even). We defined this DOS as $\Dim$ at the end of Sec.~\ref{sec_Backgrnd_NAETH}. 

Having identified the number of non-negligible terms in the $\sum_{\alpha''}$, we assess the non-negligible summand's size. The Clebsch–Gordan coefficient, dependent only on $O(1)$ numbers, is $O(1)$. The entropic factor is 
$O\big( \Dim^{-1} \big)$.
The product $f \times f \times R \times R$ is $O(1)$. Therefore, the entire summand is 
\begin{align}
   \label{eq_NAETH_scale_help1}
   O(\Dim) \times O(1) \times O \left( \Dim^{-1} \right) \times O(1)
   = O (1) .
\end{align}

Let us assemble our scaling results about Eq.~\eqref{eq_2s_help5}. By Eqs.~\eqref{eq_NAETH_scale_help0} and~\eqref{eq_NAETH_scale_help1},
\begin{align}
   \label{eq_2s_help6}
   \langle \alpha || T || \alpha \rangle
   & = O(1) + O \left( \Dim^{-1/2} \right) + O (1)
   = O(1) + O \left( \Dim^{-1/2} \right) .
\end{align}
Deterministic contributions led to the first term, whereas random ones underlie the second.

We have identified the non-Abelian ETH's indirect implication for how $\langle \alpha || T || \alpha \rangle$ scales. This indirect implication has the same form as the direct implication, Eq.~\eqref{eq_NAETH_Direct_Scale_app}. Therefore, we have proved the self-consistency of the non-Abelian ETH for block-diagonal reduced matrix elements.

\subsection{Self-consistency of the non-Abelian ETH's predictions about off--block-diagonal reduced matrix elements}
\label{app_Self_cons_Off_diag}

Recall the decomposition~\eqref{eq_combine_AB_app} of the composite spherical tensor operator $T^\KParen_q$. We must ascertain how the off--block-diagonal reduced matrix elements $\langle \alpha || T^\KParen || \alpha' \rangle$ scale. The non-Abelian ETH~\eqref{eq_NAETH} implies how directly, as in Eq.~\eqref{eq_NAETH_Direct_Scale}:
\begin{align}
   \label{eq_Self_cons_Offdiag_Direct}
   \langle \alpha || T^\KParen || \alpha' \rangle
   = 0  +  O \left( \Dim^{-1/2} \right) \times O(1) \times (1)
   = O \left( \Dim^{-1/2} \right) .
\end{align}
The nonzero term stems from a random variable.

Now, we ascertain the non-Abelian ETH's indirect implication for the element's scaling. We repeat the beginning of App.~\ref{app_Self_cons_Off_diag}, until just before Eq.~\eqref{eq_2s_help4}. That equation's off--block-diagonal analogue is
\begin{align}
   \label{eq_Self_cons_Offdiag_help1} 
   \langle \alpha || T^\KParen || \alpha' \rangle
   & = \Upsilon ( s_\alpha , s_{\alpha'} \, , s_\alpha ) \,
   \mathcal{A} \left( \mathcal{E}_{\alpha \alpha}, \mathcal{S}_{\alpha \alpha} \right) \,
   e^{- S_\th \big( \mathcal{E}_{\alpha \alpha'} \, , \, \mathcal{S}_{\alpha \alpha'} \big) / 2 } \:
   f^{(B)}_{ \nu_{\alpha \alpha' } }
   \left( \mathcal{E}_{\alpha \alpha'} \, , \mathcal{S}_{\alpha \alpha' } \, , \omega_{\alpha \alpha'} \right) \, 
   R^{(B)}_{\alpha \alpha'}
   \\ \nonumber & \quad
   + \Upsilon (s_\alpha , s_{\alpha'} \, , s_{\alpha'} ) \,
   \mathcal{B} \left( \mathcal{E}_{\alpha' \alpha'} \, , \mathcal{S}_{\alpha' \alpha'} \right) \,
   e^{- S_\th \big( \mathcal{E}_{\alpha \alpha'} \, , \, \mathcal{S}_{\alpha \alpha'} \big) / 2} \:
   f^{(A)}_{ \nu_{\alpha \alpha'} }
   \left( \mathcal{E}_{\alpha \alpha'} \, , \mathcal{S}_{\alpha \alpha'} \, , \omega_{\alpha \alpha'} \right) \,
   R^{(A)}_{\alpha \alpha'}
   \\ \nonumber & \quad
   + \sum_{\alpha''} \Upsilon (s_\alpha, s_{\alpha'} \, , s_{\alpha''} ) \,
   e^{- \big[ S_\th \big( \mathcal{E}_{\alpha \alpha''} \, , \, \mathcal{S}_{\alpha \alpha''} \big)
              + S_\th \big( \mathcal{E}_{\alpha'' \alpha'} \, , \, \mathcal{S}_{\alpha'' \alpha'} \big) 
         \big] / 2} \:
   f^{(A)}_{ \nu_{\alpha \alpha''} } \left( \mathcal{E}_{\alpha \alpha''} \, , \mathcal{S}_{\alpha \alpha''} \, , \omega_{\alpha \alpha''} \right) \,
   \\ \nonumber & \qquad \qquad \times
   f^{(B)}_{ \nu_{\alpha'' \alpha'} } \left( \mathcal{E}_{\alpha'' \alpha'} \, , \mathcal{S}_{\alpha'' \alpha'} \, , \omega_{\alpha'' \alpha'} \right) \,
   R^{(A)}_{\alpha \alpha''} \,
   R^{(B)}_{\alpha'' \alpha'} \, . 
\end{align}

We assess the terms' scalings as in App.~\ref{app_Self_cons_Off_diag}, beginning with the first two terms. $\Upsilon$ is $O(1)$ under the conditions~\eqref{eq_s_Large}--\eqref{eq_s_Large_2}, we have shown. Also $\mathcal{A}$, $\mathcal{B}$, $R^{(B)}_{\alpha \alpha'}$, and $R^{(A)}_{\alpha \alpha'}$ are $O(1)$. Each $e^{-S_\th (\ldots) / 2}$ is $O( \Dim^{-1/2} )$. Therefore, each of the first two terms in Eq.~\eqref{eq_Self_cons_Offdiag_help1} is $O( \Dim^{-1/2} )$ and contains a random contribution.

Now, we assess the final term's scaling. Again, $\Upsilon$ is $O(1)$ under the conditions~\eqref{eq_s_Large}--\eqref{eq_s_Large_2}. The $f$ functions are $O(1)$. They are non-negligible only if $| \omega_{\alpha \alpha''} |$ and $| \omega_{\alpha'' \alpha'} |$ are small---if $E_\alpha \approx E_{\alpha''} \approx E_{\alpha'}$. Similarly, since $k_1, k_2, k = O(1)$, $s_\alpha \approx s_{\alpha'} \approx s_{\alpha''} \, .$ The number of terms that meet these two conditions is approximately $\Dim = e^{ S_\th( \mathcal{E}, \mathcal{S} ) }$. Furthermore, we can approximately factor the exponential [in the final term of Eq.~\eqref{eq_Self_cons_Offdiag_help1}] outside the sum; that exponential is
$\approx e^{-S_\th ( \mathcal{E}, \mathcal{S})} 
= O (\Dim^{-1})$.
The product 
$R^{(A)}_{\alpha \alpha'} \,
   R^{(B)}_{\alpha'' \alpha'}$
is a random variable of zero mean and $O(1)$ variance. By the previous subsection's random-walk argument, therefore, the 
$\sum_{\alpha''} \Upsilon \times f \times f \times R \times R
= O (\Dim^{1/2})$. Hence the entire sum is of size
\begin{align}
   \label{eq_Self_cons_Offdiag_help2}
   O \left( \Dim^{-1} \right) \times O \left( \Dim^{1/2} \right)
   = O \left( \Dim^{-1/2} \right)
\end{align}
and originates from random variables.

Let us assemble the previous two paragraphs' conclusions about $\langle \alpha || T^\KParen || \alpha' \rangle$ when $\alpha' \neq \alpha$ [Eq.~\eqref{eq_Self_cons_Offdiag_help1}]. The reduced matrix element scales, according to the non-Abelian ETH's indirect implication, as
\begin{align}
   O \left( \Dim^{-1/2} \right) + O \left( \Dim^{-1/2} \right) + O \left( \Dim^{-1/2} \right) 
   = O \left( \Dim^{-1/2} \right) .
\end{align}
The non-Abelian ETH implies the same scaling directly [Eq.~\eqref{eq_Self_cons_Offdiag_Direct}]. Hence we have proved the self-consistency of the non-Abelian ETH's predictions about off--block-diagonal matrix elements.

\subsection{Approximation of Clebsch–Gordan coefficient}
\label{sec_CG_approx}

Let us derive the approximation~\eqref{eq_CG_approx} for the Clebsch–Gordan coefficient 
$\langle s + \nu, m+q | s, m; k, q \rangle$.
We assume that 
\begin{align}
   \label{eq_CG_approx_assume}
   s, m = O(N^\delta) , 
   \; \; \text{wherein} \; \; \delta \in (0, \, 1] , 
   \; \; \text{and} \;  \;
   \nu, k, q, s-m = O(N^0) .
\end{align}
The general formula for a Clebsch–Gordan coefficient appears as Eq.~(2.41) in~\cite{Bohm_Quantum_Book}:
\begin{align}
   \label{eq_CG_help1}
   \langle s + \nu, m + q | s, m; k, q \rangle
   & = \Big\{ [2 (s+\nu)+1]  \, (2s+\nu-k)!  \,  (k+\nu)!  \,  (k-\nu)!  \,  
                   (s+m+\nu+q)!  \,  (s-m+\nu-q)!  
                   \nonumber \\ & \qquad \; \times    
                   (s+m)!  \,  (s-m)!  \,  (k+q)!  \,  (k-q)! 
                   \, \big/ \, 
                   (2s+\nu+k+1)!  \Big\}^{1/2}
    \nonumber \\ & \quad \; \times 
    \sum_\ell \frac{ (-1)^\ell }{ \ell!  \,  (k-\nu- \ell)!  \,  (s-m- \ell)!  \,  (k+q- \ell)!  \,  (s+m+\nu-k+ \ell)!  \,  (\nu-q+ \ell)! } \, .
\end{align}
The sum runs over all integers $\ell$ for which every factorial's argument is non-negative.

To approximate the Clebsch–Gordan coefficient, we approximate factorials. If $x \gg \Delta$, then
\begin{equation}
   \label{app_large_factorial}
    (x+\Delta)! 
    = \big[  (x+\Delta)  \,  (x + \Delta - 1)  \,  \ldots  (x + 1)  \big]  \,  x!  
    = x^\Delta  \,  x!  \,  \left[1  +  O \left(x^{-1} \right)  \right] .
\end{equation}
To apply Eq.~\eqref{app_large_factorial}, we invoke the assumptions~\eqref{eq_CG_approx_assume}. For example, 
$(2s+\nu-k)! = (2s)^{\nu-k} \, (2s)!  \left[1 + O \left( N^{-\delta} \right) \right]$. In another example, if $\xi = O(1)$, then
\begin{align}
   (s + m + \xi)!
   = [2s + (s - m) + \xi]!
   = (2s)^{s - m + \xi} \, (2s)! \,
   \left[ 1 + O \left( N^{-\delta} \right) \right] .
\end{align}
We approximate all the relevant factors in Eq.~\eqref{eq_CG_help1}: at leading order,
\begin{align}
    \label{eq_CG_help2}
    \langle s+\nu, m+q|s,m; k,q\rangle 
    = (2s)^{ \frac{q-\nu}{2} }  \sum_{\ell  =  \ell_{\rm min} }^{ \ell_{\rm max} }
    \frac{ (-1)^\ell  \,  
              \sqrt{ (k+\nu)!  \,   (k-\nu)!  \,   (s-m+\nu-q)!   \,  (s-m)!  \,  (k+q)!  \, (k-q)! } }{
             \ell!  \,  (k-\nu-\ell)!   \, (s-m-\ell)!  \,  (k+q-\ell)!  \,  (\nu-q+\ell)!  \,  (2s)^\ell } \: .
\end{align}
We have defined the extreme $\ell$ values
$\ell_{\rm min} \coloneqq \max \{ 0,  \,  q-\nu \}$ and 
$\ell_{\rm max} \coloneqq \min \{ k-\nu,  \,  s-m, \,  k+q \}$.

We can approximate Eq.~\eqref{eq_CG_help2} because the summand's denominator contains a $(2s)^\ell$. This factor suppresses all terms except for the $\ell {=} \ell_{\rm min}$ one: at leading order,
\begin{equation}
    \langle s+\nu, m+q|s,m; k,q\rangle 
    = (-1)^{\ell_{\rm min}} 
    (2s)^{\frac{q-\nu}{2} - \ell_{\rm min}} 
    \frac{ \sqrt{ (k+\nu)!  \,  (k-\nu)!  \,  (s-m+\nu-q)!  \,  (s-m)!  \,  (k+q)!  \, (k-q)! }}{
             \ell_{\rm min}!  \, (k-\nu- \ell_{\rm min})!  \,  (s-m- \ell_{\rm min})!  \,  (k+q- \ell_{\rm min})!  \,  (\nu-q+ \ell_{\rm min})!} \, .
\end{equation}
We evaluate the RHS explicitly at each of $\nu > q$, $\nu = q$, and $\nu < q$:
\begin{equation}
   \label{eq_CG_help3}
    \langle s + \nu, m + q | s, m; k, q \rangle 
    = \begin{cases}
    \sqrt{ \frac{ (k+\nu)!  (s-m+\nu-q)!  (k-q)! }{
             (k-\nu)!  (s-m)!  (k+q)! } } 
   \frac{ (2s)^{\frac{q - \nu}{2} } }{ (\nu-q)! }  
   \left[ 1 + O \left( s^{-1} \right) \right] , 
   & \nu > q  \\
   1+ O \left( s^{-1}  \right) ,
   & \nu = q \\
   (-1)^{q-\nu} 
   \sqrt{ \frac{ (k-\nu)!  (k+q)!  (s-m)! }{
                     (k+\nu)!  (k-q)!  (s-m+\nu-q)! } } 
   \frac{ (2s)^{-\frac{q-\nu}{2} } }{ (q-\nu)! } 
   \left[ 1 + O \left( s^{-1} \right) \right]  ,
   &\nu < q \, .
\end{cases}
\end{equation}
This formula implies Eq.~\eqref{eq_CG_approx}.

\section{Additional numerical data about the main text's Heisenberg chain}
\label{app:data}

This appendix contains extra data about the main text's Heisenberg chain, which obeys open boundary conditions. 
First, we augment the main text's analysis of the $T^\1_0$ operator, re-analyzing the data behind Fig.~\ref{fig:Q18_Energy_gaps_GEO}. The new analysis bolsters the main-text claim that $\bra{\alpha} | T^\1 | \ket{\alpha}$ seems not unlikely to vary smoothly with $E_\alpha/N$ and $s_\alpha/N$ in the thermodynamic limit. To complement the $T^\1_0$ analysis, we calculate the properties of a spherical tensor operator $T^\2_0$. This operator acts nontrivially on two qubits, unlike the main text's $T^\1_0$. We therefore help support the generality of the main text's numerical results (Sec.~\ref{sec_results}). Also, we numerically support the expectation that $f^{(T)}_\nu$ should not depend on $q$, in accordance with the Wigner--Eckart theorem. All data in this appendix describe an 18-qubit chain, unless we specify otherwise.
 
\begin{figure}[ht]
    \centering
    \includegraphics[width=0.5\linewidth]{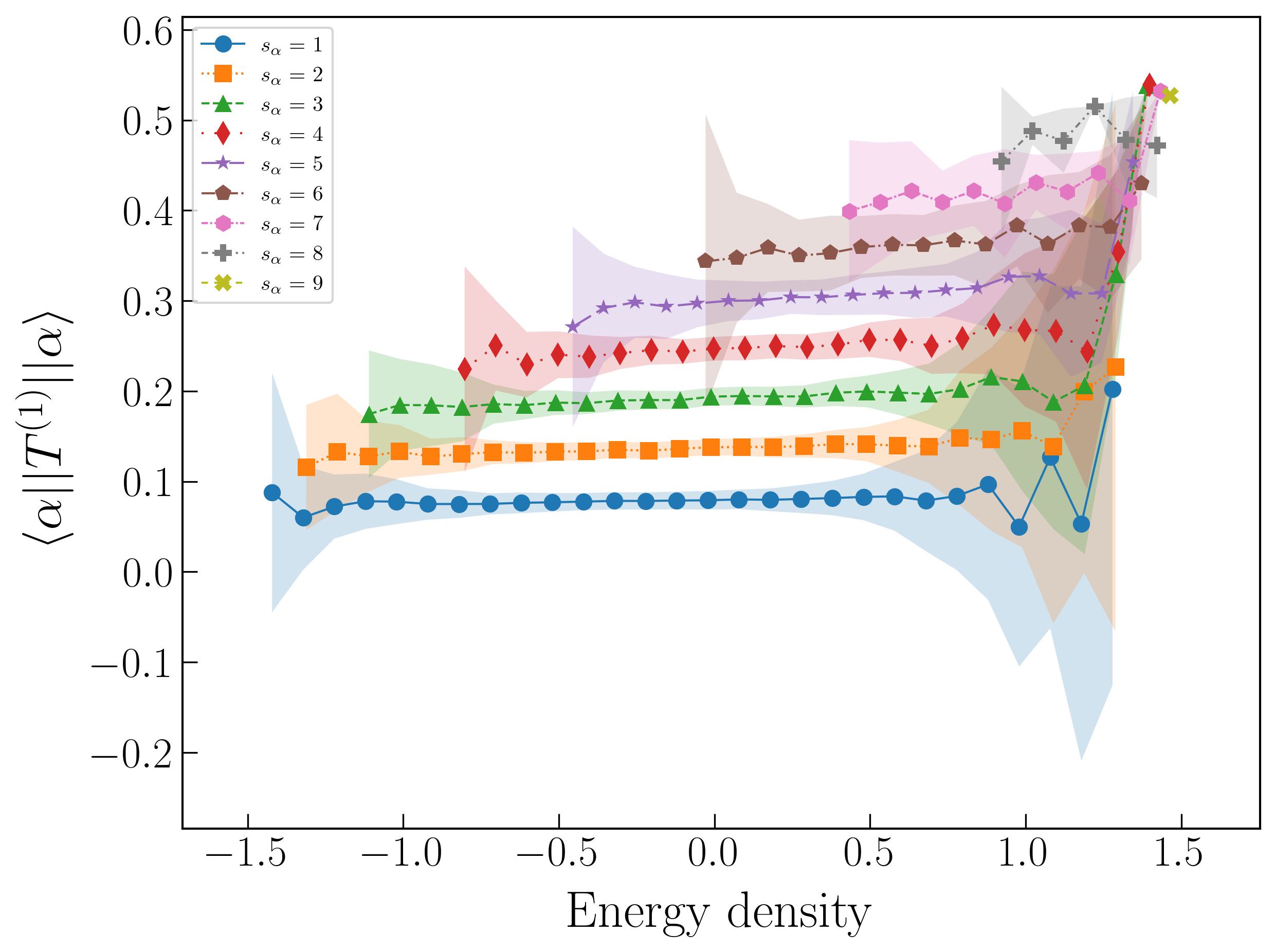}
    \caption{Average block-diagonal reduced matrix elements 
    $\langle \alpha || T^{\1} || \alpha \rangle$ versus energy density. The average is over a narrow energy-density window (of width 0.1). The elements form bands labeled by $s_\alpha$. 
    At each $s_\alpha$ value, the average varies approximately linearly with energy density, suggesting smoothness in $E_\alpha/N$. Furthermore, the average $\langle \alpha || T^{\1} || \alpha \rangle$ varies with $s_\alpha$ monotonically and nearly linearly. Hence $\langle \alpha || T^{\1} || \alpha \rangle$ seems not unlikely to vary smoothly with $E_\alpha/N$ and $s_\alpha/N$ in the thermodynamic limit, as predicted by the non-Abelian ETH's $\mathcal{T}^\KParen ( E_\alpha, s_\alpha )$ factor. The shaded regions represent standard deviations. This figure follows from the same data as Fig.~\ref{fig:Q18_Energy_gaps_GEO}.  }
    \label{fig:DiagShaded}
\end{figure}

Figure~\ref{fig:DiagShaded} highlights the block-diagonal reduced matrix elements $\langle \alpha || T^\1 || \alpha \rangle$. The figure depicts the same data as Fig.~\ref{fig:Q18_Energy_gaps_GEO}, but differently. Each marker shows the reduced matrix element, at a fixed $s_\alpha$ value, averaged over a narrow energy-density window (of width 0.1). This average $\langle \alpha || T^\1 || \alpha \rangle$ appears as a function of the energy density, $E_\alpha/N$. Lines connect adjacent markers (associated with the same $s_\alpha$ value) for readability and to emphasize the approximate smoothness with respect to $E_\alpha/N$: within each $s_\alpha$ sector, the average $\langle \alpha || T^\1 || \alpha \rangle$ varies with $E_\alpha/N$ approximately linearly. Furthermore, the standard deviations (shaded regions) are mostly small relative to the averages. (The standard deviations grow large at large $E_\alpha/N$ values. The reason is that the density of states is small here, so our sample size is small.)
In summary, the average $\langle \alpha || T^\1 || \alpha \rangle$ varies mostly smoothly with $E_\alpha/N$ at each $s_\alpha$ value. This behavior suggests that 
$\langle \alpha || T^{\1} || \alpha \rangle$ varies smoothly with $E_\alpha/N$ in the thermodynamic limit. Furthermore,
$\langle \alpha || T^{\1} || \alpha \rangle$ seems not unlikely to vary with $s_\alpha/N$ smoothly in the limit: 
the average $\langle \alpha || T^{\1} || \alpha \rangle$ depends on $s_\alpha$ monotonically and nearly linearly.
These results support the first term in the non-Abelian ETH~\eqref{eq_NAETH}. The random factor $R^{(T)}_{\alpha \alpha}$ explains the standard deviations.

\begin{figure}[ht]
    \includegraphics[width=0.5\linewidth]{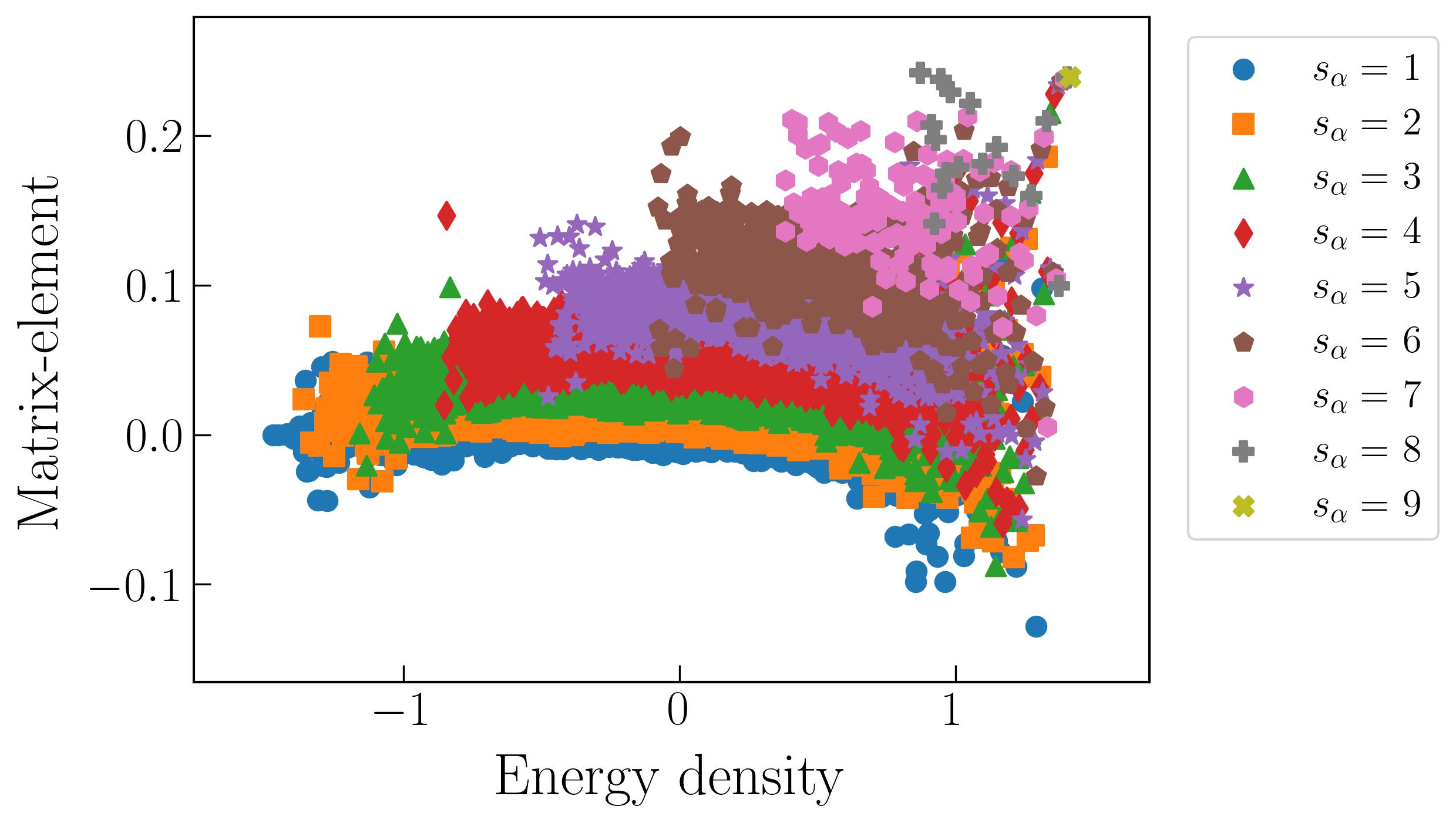}
    \caption{Block-diagonal reduced matrix elements 
    $\langle \alpha || T^\2 || \alpha \rangle$ versus energy density. 
    }
    \label{fig:Q18_T20DiagSc}
\end{figure}

\begin{figure}[ht]
    \centering
    \includegraphics[width=0.5\linewidth]{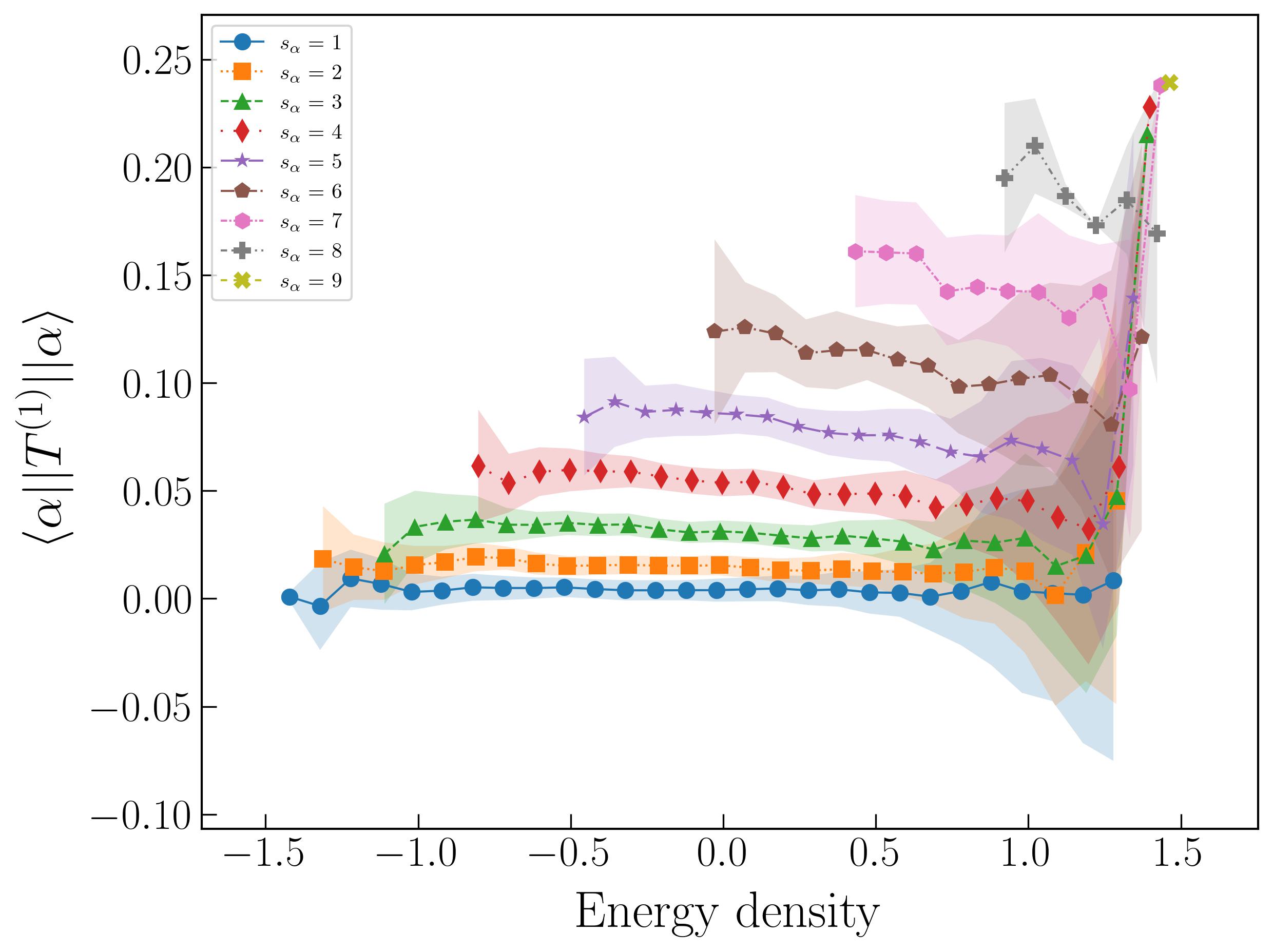}
    \caption{Average block-diagonal reduced matrix elements 
    $\langle \alpha || T^{\2} || \alpha \rangle$ versus energy density. The average is over a narrow energy-density window (of width 0.1). The elements form bands labeled by $s_\alpha$. 
The shaded regions represent standard deviations. This figure follows from the same data as Fig.~\ref{fig:Q18_T20DiagSc}. The average $\langle \alpha || T^{\2} || \alpha \rangle$ varies with $E_\alpha/N$ and $s_\alpha/N$ in a manner suggestive of smoothness of $\mathcal{T}^\KParen ( E_\alpha, s_\alpha )$.}
    \label{fig:DiagShadedT20}
\end{figure}

\begin{figure}[h]
\centering
\begin{subfigure}{0.4\textwidth}
\centering
\includegraphics[width=\linewidth]{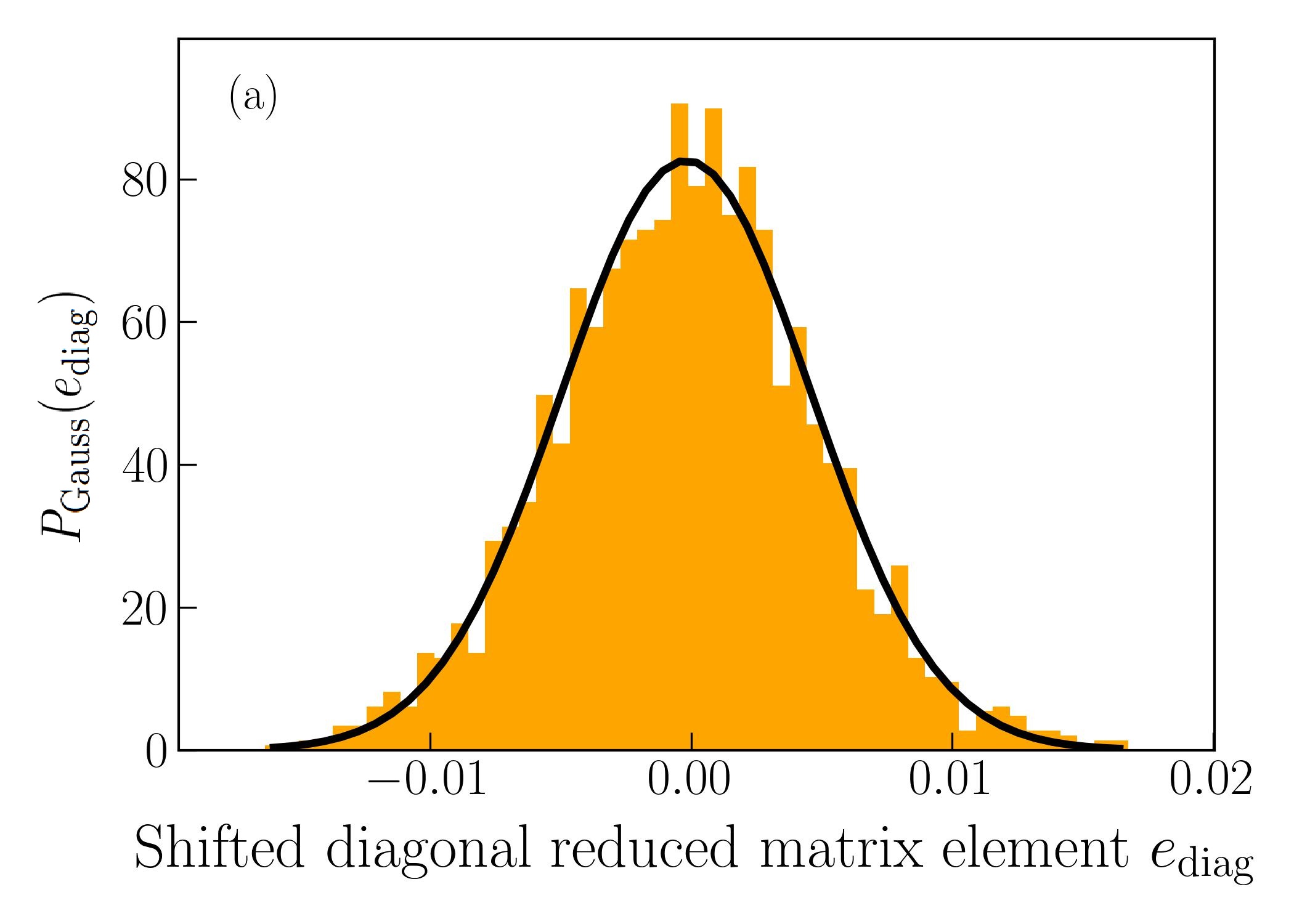}
\end{subfigure}
\begin{subfigure}{0.4\textwidth}
\centering
\includegraphics[width=\linewidth]{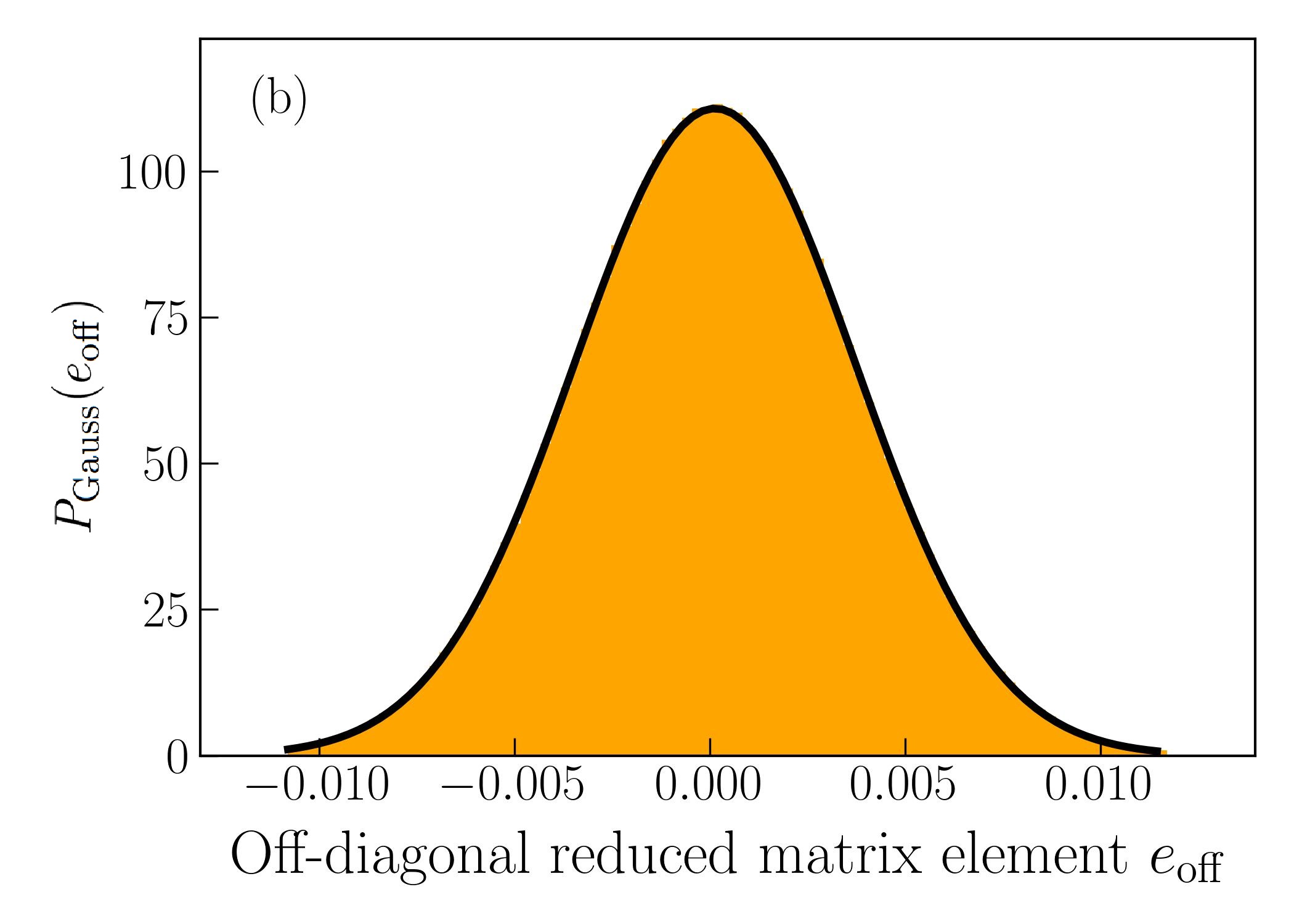}
\end{subfigure}
\caption{Histograms of reduced matrix elements. The eigenenergies $E_\alpha$ come from a narrow energy-density window (of width 0.1). The energy eigenstates used come from the representative $s_\alpha {=}3$ band depicted in Fig.~\ref{fig:Q18_T20DiagSc}. The black curves represent Gaussian distributions. (a) Histogram of block-diagonal reduced matrix elements, shifted so that their averages vanish. The Gaussian distribution's fit has a coefficient of determination $R^2=0.981$ (b) Histogram of off--block-diagonal reduced matrix elements 
$\langle \alpha || T^\2 || \alpha' \rangle$. The Gaussian distribution's fit has a coefficient of determination $R^2=0.999$.
}
\label{fig_Histograms_App}
\end{figure}

\begin{figure}[h]
\centering
\begin{subfigure}{0.48\textwidth}
\centering
\includegraphics[width=\textwidth]{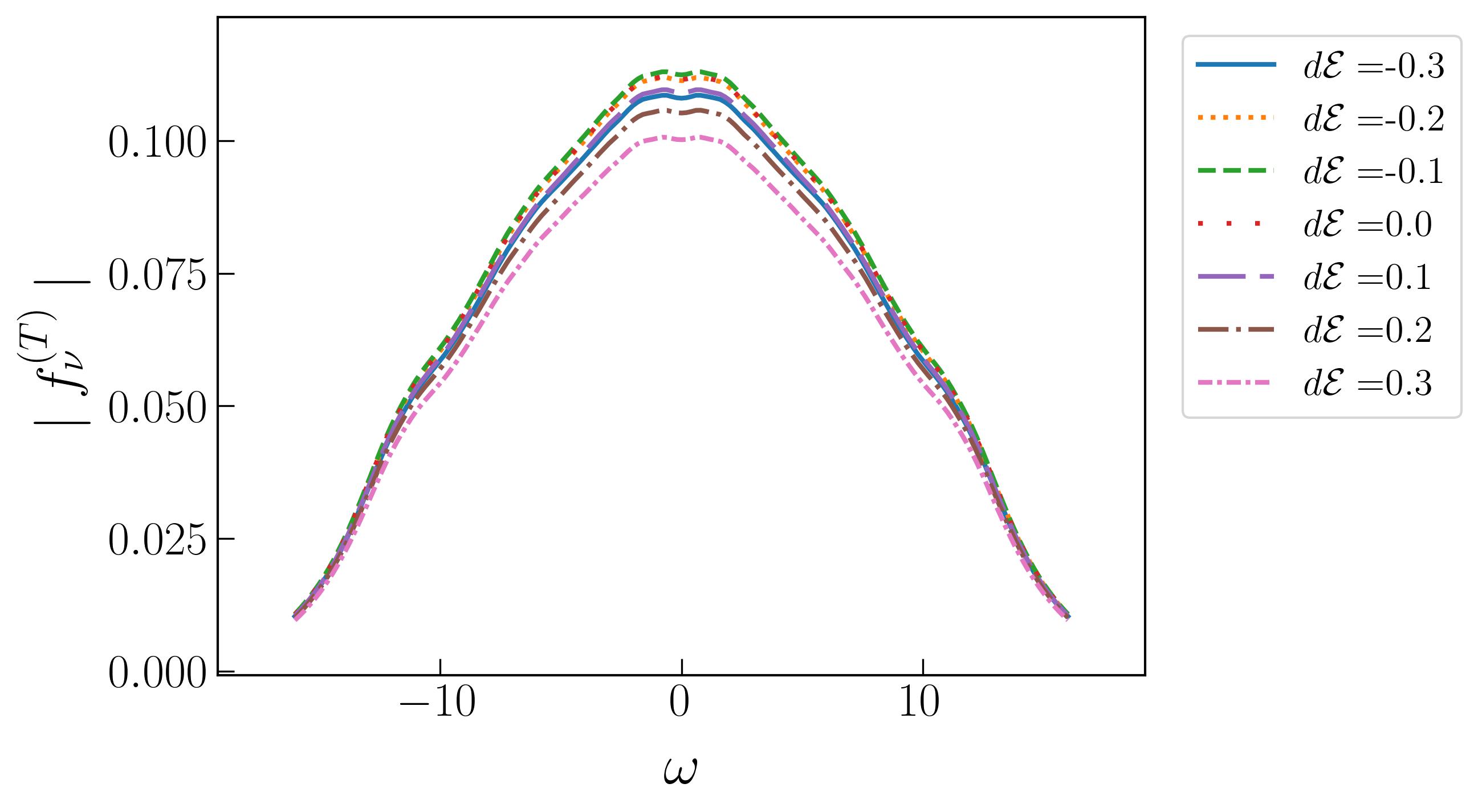}
\end{subfigure}
\begin{subfigure}{0.48\textwidth}
\centering
\includegraphics[width=\textwidth]{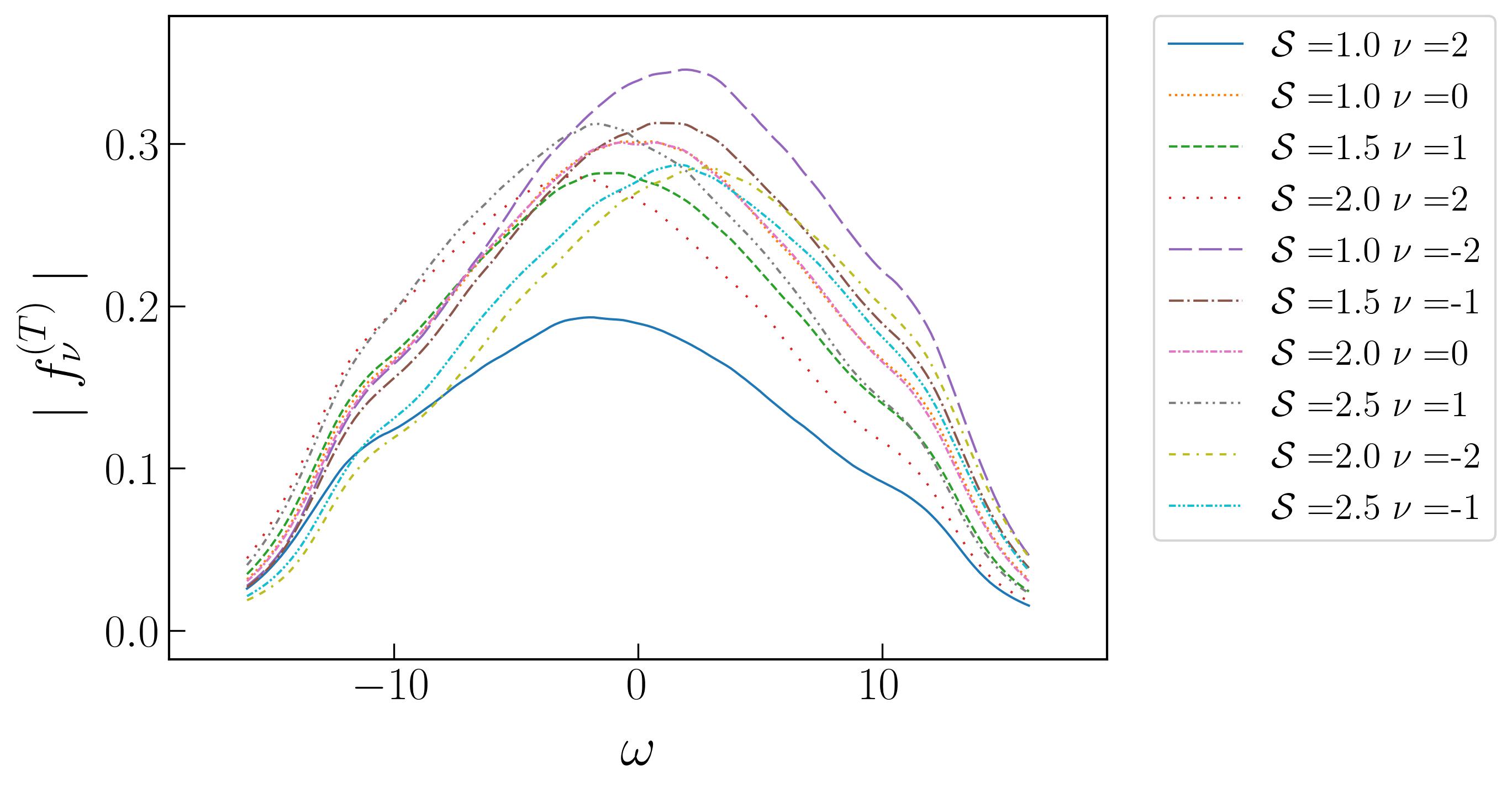}
\end{subfigure}
\caption{Magnitude of $f^{(T)}_\nu$, as a function of $\omega$, with $\mathcal{S}$ or $\mathcal{E}$ fixed. 
(a) Magnitude of 
$f^{(T)}_{\nu=0}\left( \mathcal{E}, \mathcal{S}{=}3, \omega \right)$. Different curves correspond to different $\mathcal{E}$ values. (b) Magnitude of $f^{(T)}_\nu \left( \mathcal{E} {=} 0, \mathcal{S}, \omega \right)$. Different curves correspond to different $\mathcal{S}$ and $\nu$ values.}
\label{fig_f_app}
\end{figure}

\begin{figure}[ht]
    \centering
    \includegraphics[width=0.5\linewidth]{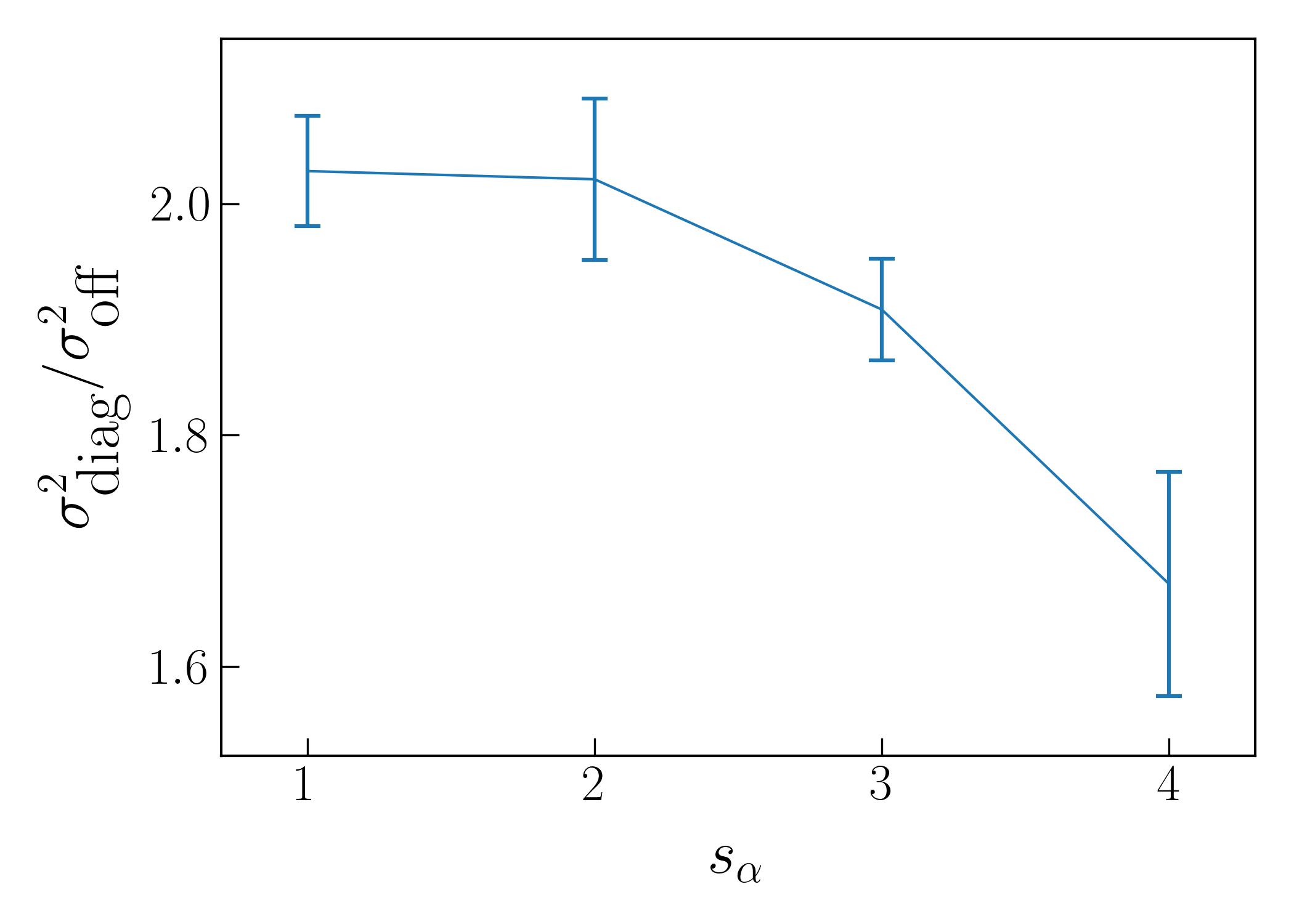}
    \caption{Average of $\sigma_\text{diag}^2 / \sigma_\text{off}^2$, as a function of the spin quantum number $s_\alpha$, for the operator $T_0^{(2)}$. The average is over many narrow energy-density windows. 
    For each $s_\alpha$, the average ratio is close to the value predicted with random-matrix theory, 2, and average of $\sigma_\text{diag}^2 / \sigma_\text{off}^2$ over $s_\alpha$ = $1.91 \pm 0.14$.} 
    \label{fig:varratios_T20}
\end{figure}

Figures~\ref{fig:Q18_T20DiagSc}--\ref{fig:varratios_T20} echo figures in Sec.~\ref{sec_results} but describe the spherical tensor operator
\begin{align}
   T^\2_0 
   = \frac{1}{\sqrt{24}} \left (  \sigma_z^{\left ( \lceil N/2 \rceil \right )} \sigma_z^{\left ( \lceil N/2 \rceil +1 \right )} - \sigma_+^{\left ( \lceil N/2 \rceil \right )} \sigma_-^{\left ( \lceil N/2 \rceil +1 \right )}   - \sigma_-^{\left ( \lceil N/2 \rceil \right )} \sigma_+^{\left ( \lceil N/2 \rceil +1 \right )} \right ) .
\end{align}
Figure~\ref{fig:Q18_T20DiagSc} shows the block-diagonal reduced matrix elements
$\langle \alpha || T^\2 || \alpha \rangle$ plotted against energy density. As in the main text, the markers form bands labeled by $s_\alpha$. Figure~\ref{fig_Histograms_App} shows histograms formed from the $s_\alpha{=}3$ eigenspace.
In Fig.~\ref{fig_Histograms_App}(a), we histogram the block-diagonal matrix elements, shifted so that their average vanishes. Each shifted matrix element is expected to equal the non-Abelian ETH's second term: 
$\langle \alpha || T^\2 || \alpha \rangle
- \mathcal{T}^\2 ( \mathcal{E}, \mathcal{S} )$.
In Fig.~\ref{fig_Histograms_App}(b), we histogram the off--block-diagonal reduced matrix elements 
$\langle \alpha || T^\KParen || \alpha' \rangle$. Gaussian distributions fit both histograms, consistently with the non-Abelian ETH's $R^{(T)}_{\alpha \alpha'} \, .$ Figure~\ref{fig_f_app} shows the magnitude 
$| f^{(T)}_\nu \left( \mathcal{E}, \mathcal{S}, \omega \right) |$ as its arguments vary. The results suggest that $f^{(T)}_\nu$ depends on the arguments smoothly in the thermodynamic limit. Figure~\ref{fig:varratios_T20} shows the average ratio of the block-diagonal elements' variance, $\sigma^2_\text{diag}$, to the off--block-diagonal elements', $\sigma^2_\text{off}$ as a function of $s_\alpha$. 

We now support the expectation that the non-Abelian ETH---including every factor in its RHS---does not depend on $q$. This independence follows from the Wigner--Eckart theorem [Eq.~\eqref{eq_WE_thm}]. $q$ affects computations' tractability, however: consider representing any $T^\KParen_q$ relative to the basis $\{ \ket{\alpha, m} \}$. The matrix is symmetric if $q=0$, simplifying calculations and saving memory. 
We therefore calculated Figs.~\ref{fig:Q18_Energy_gaps_GEO}--\ref{fig:varratios_T10} from $T^\1_0$ and calculated Figs.~\ref{fig:Q18_T20DiagSc}--\ref{fig_f_app} from $T^\2_0$, for convenience. However, we also calculated the reduced matrix elements of
\begin{align}
   T^\1_1 
   = -  \frac{1}{\sqrt{2}} \sigma_+^{\left ( \lceil N/2 \rceil \right )}
   \quad \text{and} \quad
   T^\2_2 
   = \frac{1}{2}  \sigma_+^{\left ( \lceil N/2 \rceil \right )} \sigma_+^{\left ( \lceil N/2 \rceil +1 \right )} \, .
\end{align}
We confirmed that these elements---and so their statistics and $f^{(T)}_\nu$ functions---do not depend on $q$, to within numerical precision.

\section{Second numerical example: Larger 1D Heisenberg model with periodic boundary conditions and extra symmetries}
\label{app_JDN_Num}

This appendix concerns a model alternative to the main text's 1D Heisenberg chain with open boundary conditions and only SU(2) symmetry. Here, we analyze a 1D Heisenberg chain subject to periodic boundary conditions. The chain also has translational and spatial-reflection symmetries. The extra symmetries enable us to model a larger system than in the main text---26, rather than 18, qubits. This appendix's results qualitatively resemble the main text's, supporting them and suggesting their generality.

Our 1D Heisenberg model contains an even number $N = 26$ of qubits. The chain obeys periodic boundary conditions: 
$\vec{\sigma}^\JParen 
= \vec{\sigma}^{(j + N)}$.
The Hamiltonian has nearest-neighbor and (to break integrability) next-nearest-neighbor couplings:
\begin{equation}\label{H_pbc}
    H_{\rm PBC} 
    = J \sum_{j=1}^N \left( \vec{\sigma}^{(j)} \cdot 
        \vec{\sigma}^{(j+1)} +
    \vec{\sigma}^{(j)} \cdot \vec{\sigma}^{(j+2)} \right) .
\end{equation}
We set $J$ to 1. $H_{\rm PBC}$ remains invariant under transformations under four relevant types:
\begin{enumerate}
   \item Identical arbitrary rotations of all the qubits [elements of SU(2)].
   \item \label{sym_Trans}
   Translations (generated by the shift operator $\Delta$).
   \item Spatial inversion (or a reflection about the chain's midpoint. We denote by $R$ the operator that generates the reflection).
   \item \label{sym_Inv}
   Spin inversion (conjugation by $X \coloneqq \bigotimes_{j=1}^N \sigma_x^\JParen$).
\end{enumerate}

To facilitate calculations, we focus on a subspace of the $2^N$-dimensional global Hilbert space. The symmetry operators of~\ref{sym_Trans}--\ref{sym_Inv}, as well as $S_z$, share an eigenspace called the \emph{maximum-symmetry sector} (MSS)~\cite{Jung_20_MSS_ExactDiagonalization,JaeDongXXZ}. This eigenspace corresponds to the eigenvalue 1 of $\Delta$, $R$, and $X$, as well as to the eigenvalue 0 of $\sigma_z^\tot$. Hamiltonian eigenstates 
$\{ | \alpha, m{=}0\rangle\}$
span this sector. Similarly to how $m=0$ throughout the MSS, $s_\alpha$ assumes only the values $N/2, N/2-2, \ldots$ 

\begin{figure}
    \includegraphics[width=0.75\linewidth]{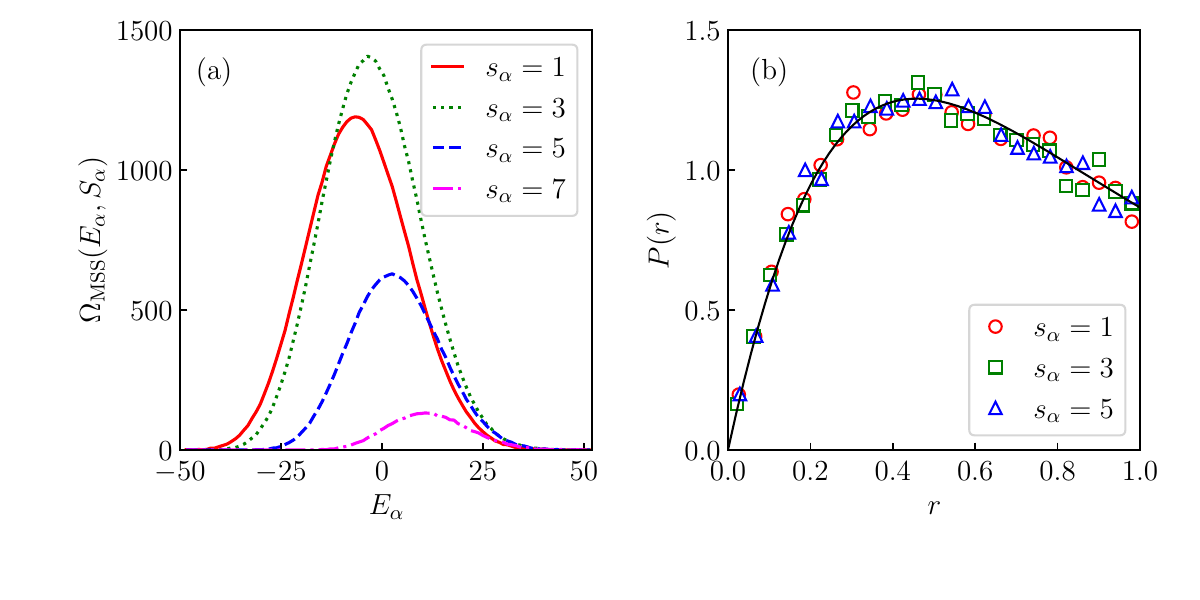}
    \caption{(a) Density of states, $\Omega_\MSS(E_\alpha,s_\alpha)$. 
    (b) Probability density $P(r)$ of the possible values $r$ of the minimal-gap ratio. $P(r)$ agrees well with the random-matrix-theory prediction $P_\GOE (r)$. The $s_\alpha{=}7$ data deviate noticeably, coming from an eigenspace much smaller than the other $\vec{S}^2$ eigenspaces.}
\label{fig:MSS_DOS_GRatio}
\end{figure}

Before testing the non-Abelian ETH, we confirm that $H_{\rm PBC}$ is nonintegrable and so should obey an ETH statement. We calculate gap-ratio statistics, as in Sec.~\ref{sec_results}, using the DOS. Denote by $\Omega_\MSS(E_\alpha, s_\alpha)$ the DOS within the $s_\alpha$ eigenspace. Figure~\ref{fig:MSS_DOS_GRatio}(a) depicts $\Omega_\MSS(E_\alpha, s_\alpha)$ as a function of the eigenenergy $E_\alpha$. The plot qualitatively resembles the 18-qubit analogue, Fig.~\ref{fig:Q18_DOS}. From the $E_\alpha$s, we calculate the minimal gap ratios
$\min\left\{ \frac{E_{n+1}-E_n}{E_n-E_{n-1}}, \frac{E_n-E_{n-1}}{E_{n+1}-E_n} \right\} \, ,$ 
as in Sec.~\ref{sec_results}. The probability density of the possible minimal-gap-ratio values $r$ appears in Fig.~\ref{fig:MSS_DOS_GRatio}(b). Different marker colors (and marker shapes) correspond to different $\vec{S}^2$ eigenspaces.
We constructed the probability density from the window, centered on the energy band's middle, that contains half of the $E_\alpha$s. The probability density agrees well with the random-matrix-theory prediction $P_\GOE(r) = \frac{27}{4}\frac{r(1+r)}{(1+r+r^2)^{5/2}} \, .$ Hence $H_{\rm PBC}$ is nonintegrable within each $\vec{S}^2$ eigenspace.

Having established the Hamiltonian's nonintegrability, we test the non-Abelian ETH. We analyze the spherical tensor operators
\begin{align}
        & T^{(0)}_0 = -\frac{1}{12N} \sum_{j=1}^N
        \left( 2 \sigma_+^{(j)}
            \sigma_-^{(j+1)} + 2 \sigma_-^{(j)}\sigma_+^{(j+1)} +
        \sigma_z^{(j)} \sigma_z^{(j+1)}\right) 
        \quad \text{and} \\
        & T^{(2)}_0 = \frac{1}{\sqrt{24}N} \sum_{j=1}^N \left( \sigma_z^{(j)}
            \sigma_z^{(j+1)} - \sigma_+^{(j)}
            \sigma_-^{(j+1)} - \sigma_-^{(j)}\sigma_+^{(j+1)} \right) .
\end{align}
We calculate these operators' reduced matrix elements from the Wigner--Eckart theorem [Eq.~\eqref{eq_WE_thm}]:
\begin{align}
    \langle \alpha \| T^{(k)} \|\alpha'\rangle 
    = \frac{\langle \alpha
        ,s_\alpha, m=0 | T^{(k)}_0 | \alpha', s_{\alpha'},m=0\rangle}
        {\langle s_\alpha,0| s_{\alpha'},0; k,0 \rangle} \, .
\end{align}

\begin{figure}
   \includegraphics[width=0.75\columnwidth]{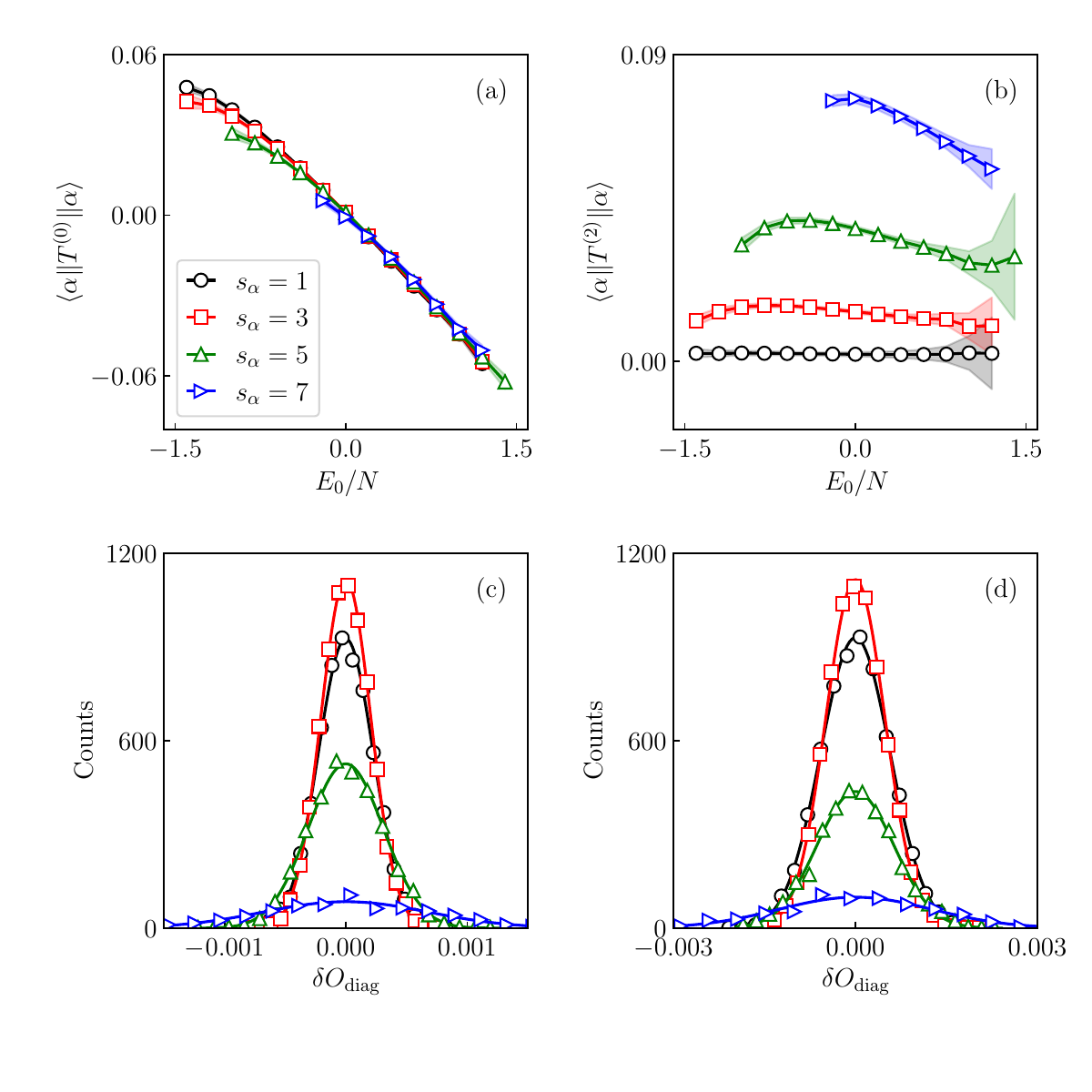}
    \caption{Statistics of block-diagonal reduced matrix elements
    $\langle \alpha || T^\KParen || \alpha \rangle$. 
    (a, b) The average elements of $T^{(0)}_0$ appear in (a); and those of $T^{(2)}_0$, in (b). We calculated each data point from a running average within an energy window, centered at $E_0$, of width $\Delta E=0.2N$. $E_0$ varies along the $x$-axis. Each shaded area's width equals twice the standard deviation. 
    (c, d) Histograms of the block-diagonal reduced matrix elements' differences $\delta O_{\rm diag}$ from their averages. We present data about $T^\0_0$ in (c) and data about $T^\2_0$ in (d). The text describes the energy windows whose statistics are depicted.
    Gaussian functions (solid curves) match the histograms, as predicted by the non-Abelian ETH's $R^{(T)}_{\alpha \alpha}$.}
    \label{fig:MSS_Diagonal}
\end{figure}

Figures~\ref{fig:MSS_Diagonal}(a) and (b) show averages and standard deviations of the block-diagonal elements, $\langle \alpha \| T^{(k)}\|\alpha\rangle$. Different marker colors (and marker shapes) correspond to different $\vec{S}^2$ eigenspaces. Each data point represents an average formed from the energies $E_\alpha$ in the window, centered on $E_0$, of width $\Delta E = 0.2N$:
$|E_\alpha-E_0|\leq \Delta E/2$.
The center $E_0$ runs over the spectrum as one progresses across the $x$-axis. The average is expected to be $\mathcal{T}^{(k)}(E_\alpha, s_\alpha)$. It varies smoothly with the energy density, as expected. The standard deviations (shaded regions) are much smaller than the averages, except near the spectrum's edges: spectra's edges are generally not expected necessarily to obey ETH statements.

We now model the fluctuations behind the standard deviations. To do so, we calculate each block-diagonal reduced matrix element's difference $\delta O_{\rm diag}$ from its average. (The average is over an energy window, specified below.) If the non-Abelian ETH is correct, this difference equals
$\langle \alpha || T^\KParen || \alpha \rangle
- \mathcal{T}^\KParen ( \mathcal{E}, \mathcal{S} )$.
We histogram the $\delta O_{\rm diag}$ of $T^\0$ in Fig.~\ref{fig:MSS_Diagonal}(c) and that of $T^\2$ in Fig.~\ref{fig:MSS_Diagonal}(d). These figures rely on different energy windows than Figs.~\ref{fig:MSS_Diagonal}(a) and (b): denote by $E_0(s_\alpha)$ the energy at which the DOS [Fig.~\ref{fig:MSS_DOS_GRatio}(a)] maximizes within the $s_\alpha$ subspace. We center the energy window at $E_0(s_\alpha)$, for each $s_\alpha$. This choice yields better statistics than energy windows that contain fewer eigenstates. Gaussian distributions fit both histograms well, consistently with the non-Abelian ETH's $R^{(T)}_{\alpha \alpha'} \, .$

\begin{figure}
    \includegraphics[width=0.75\columnwidth]{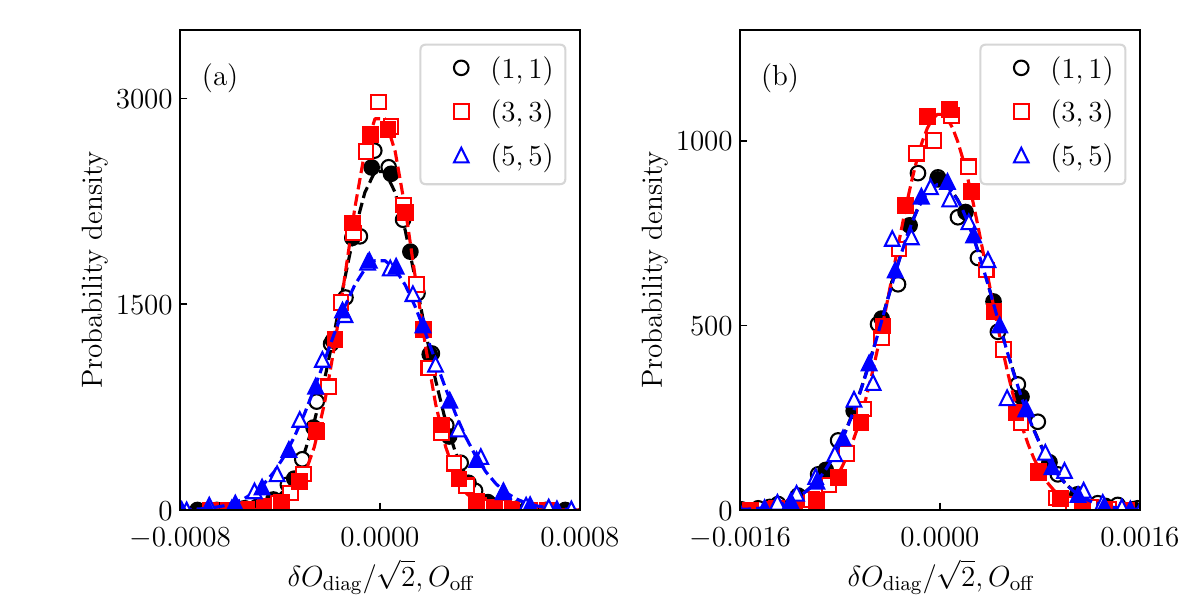}
    \caption{Probability densities of reduced matrix elements. Empty symbols represent shifted, rescaled block-diagonal elements $\delta O_{\rm diag}/\sqrt{2}$; filled symbols represent off--block-diagonal elements $O_{\rm off}$. We present data about $T^\0_0$ in (a) and data about $T^\2_0$ in (b). Different marker colors (and different marker shapes) correspond to different $s_\alpha {=} s_{\alpha'}$ values. The text describes the energy window used. Gaussian distributions (dashed curves) fit all the probability densities, regardless of $T^\KParen_q$ and $s_\alpha {=} s_{\alpha'} \, .$ Furthermore, the distributions overlap significantly, supporting the non-Abelian ETH quantitatively. }
\label{fig:diag_offdiag_T0_T2}
    \end{figure}

Finally, we calculate distributions over the possible values of the block-diagonal and off--block-diagonal reduced matrix elements $\langle \alpha \| T^{(k)} \|\alpha'\rangle$. We focus on $s_{\alpha'} = s_\alpha$, such that the block-diagonal elements are nonzero. The matrix elements are defined by the energy eigenstates ($\ket{E_\alpha}$ and $\ket{E_{\alpha'}}$) whose energies ($E_\alpha$ and $E_{\alpha'}$) lie within an energy window, centered at $E_0 = 0$, of width $\Delta E = 0.1N$: 
$E_\alpha, E_{\alpha'} \in (E_0 - \Delta E/2 , \, E_0 + \Delta E/2)$. Figure~\ref{fig:diag_offdiag_T0_T2}(a) shows data about $T^\0_0$; and Fig.~\ref{fig:diag_offdiag_T0_T2}(b), data about $T^\2_0$. Empty markers represent shifted, rescaled block-diagonal elements $\delta O_{\rm diag} / \sqrt{2}$ (we justify the rescaling below); and filled markers represent off--block-diagonal elements $O_{\rm off}$. Different marker colors (and marker shapes) represent different $s_\alpha {=} s_{\alpha'}$ values. Gaussian distributions model all the histograms well, irrespectively of $T^\KParen_q$ and $s_\alpha {=} s_{\alpha'} \, .$ Because we multiply $\delta O_{\rm diag}$ by $1/\sqrt{2}$, the block-diagonal--element and off--block-diagonal--element distributions overlap substantially. This overlap indicates that the block-diagonal elements' variance equals twice the off--block-diagonal elements' variance. This result quantitatively supports the non-Abelian ETH (see Sec.~\ref{sec_results}, especially Fig.~\ref{fig:varratios_T10}). Each variance equals
\begin{align}
   e^{-S_\th (\mathcal{E}{=}E_0, \, \mathcal{S}{=}s_\alpha)} \,
   \left\lvert f^{T}_{\nu=0}
   (\mathcal{E}{=}E_0, \mathcal{S}{=}s_\alpha, \omega{=}0) \right\rvert^2 \, .
\end{align}

\end{appendices}

\bibliography{main.bib}
\end{document}